%% file: main-650.tex
\documentclass[a4paper,11pt]{article}

\usepackage[english]{babel}
\usepackage{jheppub}

\usepackage{graphicx}
\usepackage{cancel}
\usepackage{MnSymbol}
\usepackage{bbold}
\usepackage{amsmath,amssymb,amsfonts}
\usepackage{slashed}
\usepackage{caption}
\usepackage{appendix}
\usepackage{hyperref} 
\usepackage{multirow}
\hypersetup{colorlinks=true, linkcolor=black, urlcolor=blue}
\usepackage{comment}
\usepackage{float}
\usepackage{mathtools}

\usepackage{amstext} 
\usepackage{array}   
\newcolumntype{L}{>{$}l<{$}} 

\usepackage{booktabs}

\usepackage{geometry}
\usepackage{color}
\definecolor{gray}{rgb}{0.5,0.5,0.5}

\def\det{\mbox{det}\,}

\def\SU{\mathrm{SU}}
\def\SO{\mathrm{SO}}
\def\UU{\mathrm{U}(1)}

\newcommand\lsim{\mathrel{\rlap{\lower4pt\hbox{\hskip1pt$\sim$}}
    \raise1pt\hbox{$<$}}}
\newcommand\gsim{\mathrel{\rlap{\lower4pt\hbox{\hskip1pt$\sim$}}
    \raise1pt\hbox{$>$}}}

\newcommand{\beq}{\begin{equation}}
\newcommand{\eeq}{\end{equation}}
\newcommand{\bea}{\begin{eqnarray}}
\newcommand{\eea}{\end{eqnarray}}
\newcommand{\bem}{\begin{pmatrix}}
\newcommand{\eem}{\end{pmatrix}}

\newcommand{\bet}{\begin{itemize}}
\newcommand{\eet}{\end{itemize}}
\newcommand{\ben}{\begin{enumerate}}
\newcommand{\een}{\end{enumerate}}

\def\TRANSPOSE{T}

\def\EMPH{\underline}

\definecolor{myGreen}{rgb}{0.39, 0.65, 0.46}
\definecolor{myBlue}{rgb}{0.17, 0.26, 0.65}
\definecolor{myRed}{rgb}{0.85, 0.0, 0.0}

\makeatletter
\gdef\@fpheader{}
\makeatother

\begin{document}

\title{Trinification from $\mathbf{E_{6}}$ symmetry breaking}

\author[a]{K.S.~Babu,}
\author[b]{Borut Bajc,}
\author[c]{Vasja Susič}

\affiliation[a]{Department of Physics, Oklahoma State University, Stillwater, OK, 74078, USA}
\affiliation[b]{Jožef Stefan Institute, Jamova cesta 39, 1000 Ljubljana, Slovenia}
\affiliation[c]{Institute of Particle and Nuclear Physics,
    Faculty of Mathematics and Physics, Charles University,\\ V Hole\v{s}ovi\v{c}k\'{a}ch 2, 180 00 Prague 8, Czech Republic}

\emailAdd{babu@okstate.edu}
\emailAdd{borut.bajc@ijs.si}
\emailAdd{susic@ipnp.mff.cuni.cz}

\abstract{
In the context of $\mathrm{E}_{6}$ Grand Unified Theories (GUTs), an intriguing possibility for symmetry breaking to the Standard Model (SM) group involves an intermediate stage characterized by either $\mathrm{SU}(3)\times\mathrm{SU}(3)\times\mathrm{SU}(3)$ (trinification) or $\mathrm{SU}(6)\times\mathrm{SU}(2)$. The more common choices of $\mathrm{SU(5)}$ and $\mathrm{SO}(10)$ GUT symmetry groups do not offer such breaking chains. We argue that the presence of a real (rank $2$ tensor) representation $\mathbf{650}$ of $\mathrm{E}_{6}$ in the scalar sector is the minimal and likely only reasonable possibility to obtain one of the novel intermediate stages. We analyze the renormalizable scalar potential of a single copy of the $\mathbf{650}$ and find vacuum solutions that support regularly embedded subgroups  $\mathrm{SU}(3)\times\mathrm{SU}(3)\times\mathrm{SU}(3)$, $\mathrm{SU}(6)\times\mathrm{SU}(2)$, and $\mathrm{SO}(10)\times\mathrm{U}(1)$, as well as specially embedded subgroups $\mathrm{F}_{4}$ and $\mathrm{SU}(3)\times\mathrm{G}_{2}$ that do not contain the SM gauge symmetry. We show that for a suitable choice of parameters, each of the regular cases can be obtained as the lowest among the analyzed minima in the potential. 
}

\maketitle

\section{Introduction}

An intriguing possibility of enlarging the gauge interactions of the Standard Model (SM) is to embed them together in a Grand Unified Theory (GUT)~\cite{Georgi:1974sy}. For such a GUT to be viable and consistent, its symmetry group must be a simple Lie group, which has the SM group $\SU(3)\times\SU(2)\times\UU$ as its subgroup, and must admit complex representations (so as to produce a chiral theory). Only the groups $\SU(N+3)$, $\SO(4N+2)$ and $\mathrm{E}_{6}$ satisfy these criteria, where $N\geq 2$ and the minimal cases in the unitary and orthogonal families are the familiar $\SU(5)$ and $\SO(10)$ groups. The focus of this paper is the third possibility, viz., $\mathrm{E}_{6}$.

The exceptional Lie group $\mathrm{E}_{6}$ contains $\SU(5)$ and $\SO(10)$ as its subgroups and is thus the largest of the usual GUT symmetry possibilities, as recognized long ago~\cite{Gursey:1975ki}.  Unlike $\SU(5)$ and $\SO(10)$, the group $\mathrm{E}_{6}$ also has the trinification group $\SU(3)\times\SU(3)\times\SU(3)$ and $\SU(6)\times\SU(2)$ as its subgroups (we shall sometimes denote them as $G_{333}$ and $G_{62}$ for simplicity), and thus allows breaking routes to the SM not present in the other two (minimal) GUT symmetry groups. Although 
$\mathrm{E}_{6}$ also has other phenomenologically peculiar features, such as 
matter unification of each generation joined by $2$ sterile neutrinos and vector-like exotics in the fundamental representation $\mathbf{27}$, it is the exotic intermediate symmetries that $\mathrm{E}_{6}$ can break into that are of interest to us in this paper.

A crucial role in obtaining $G_{333}$ or $G_{62}$ at the intermediate stage is played by the representation $\mathbf{650}$, since it is the simplest and seemingly only realistic\footnote{ 
    Consider an $\mathrm{E}_{6}$ 4D Yang-Mills theory with a single representation of real scalars $\mathbf{R}$, which contains $G_{333}$ or $G_{62}$ singlets. The one-loop RGE for the unified gauge coupling is $dg/dt=\beta g^3/(16\pi^2)$. The two simplest candidates for $\mathbf{R}$ are the representations  $\mathbf{650}$ and $\mathbf{2430}$, yielding $\beta=-19$ and $\beta=+91$, respectively. The second case leads to a fast Landau pole, hence the strong preference for $\mathbf{650}$ in model building.
}
representation that contains a SM-singlet of either, see e.g.~\cite{Slansky:1981yr,Feger:2019tvk}. Any reasonable GUT model utilizing the $G_{333}$  or $G_{62}$ intermediate symmetries will thus require at least one scalar representation $\mathbf{650}$ in its symmetry-breaking sector. 

In this paper, we determine the (non-supersymmetric) scalar potential of a single representation $\mathbf{650}$ and study its minima. This effort can be seen as a first step in analyzing the class of $\mathrm{E}_{6}$ models that aspires to use an exotic breaking route to the SM. We limit ourselves to the analysis of the first breaking stage, which should be followed by at least one more before reaching the SM. Also, although other representations are required for a realistic model,\footnote{We note here that adding scalar representations $\mathbf{27}\oplus\mathbf{351}'$ to the $\mathbf{650}$ seems especially promising to us, since their addition is well suited to describing a realistic Yukawa sector, that reduces to the $\mathbf{10}\oplus\mathbf{126}$ case in $\SO(10)$ \cite{Babu:1992ia,Bajc:2005zf,Joshipura:2011nn,Ohlsson:2019sja} in the limit of $\SO(10)$-spinors in the $\mathrm{E}_{6}$ scalar representations not acquiring VEVs (deducible from~\cite{Babu:2015psa}). 
}
we do not commit to any particular fixed scalar sector beyond the presence of one copy of the $\mathbf{650}$.

Since the inclusions $\mathbf{75}\subset\mathbf{210}\subset\mathbf{650}$ describe the largest irrep subparts with respect to the subgroup chain $\SU(5)\subset\SO(10)\subset\mathrm{E}_{6}$, the study in this paper can be viewed as an $\mathrm{E}_{6}$ analogue to the vacuum analysis studies of the $\mathbf{75}$ in $\SU(5)$ GUT~\cite{Koh:1982ji,Hubsch:1984pg,Hubsch:1984qi,Cummins:1985vg} or $\mathbf{210}$ in $\SO(10)$~\cite{Chang:1984qr,Chang:1985zq,Chang:1986be,He:1989rb}. 
Within $\mathrm{E}_{6}$ GUT, the representation $\mathbf{650}$ has been considered before~\cite{Lonsdale}, albeit with a partial set of invariants.

We organize the paper as follows: we present some group theory preliminaries required for describing the representation $\mathbf{650}$ in Section~\ref{sec:preliminaries}, write down and analyze the scalar potential using one copy of this representation in Section~\ref{sec:potential}, and then conclude in Section~\ref{sec:conclusions}. We also provide a set of appendices, to which further technical details have been relegated.

\section{Group Theory preliminaries of $\mathrm{E}_{6}$ \label{sec:preliminaries}}

\subsection{Generators and representations of $\mathrm{E}_{6}$ \label{sec:e6-generators}}

The exceptional group $\mathrm{E}_{6}$ is simple, contains the SM group $\SU(3)_C\times\SU(2)_L\times\mathrm{U}(1)_{Y}\equiv G_{321}$ and has complex representations, and is thus a suitable candidate for the unification group in GUT.

The $\mathrm{E}_{6}$ group has $78$ generators and is of rank $6$, see e.g.~\cite{Slansky:1981yr,Feger:2019tvk} for group theory resources. The most convenient way of considering the generators of $\mathrm{E}_6$ is through the language of the trinification subgroup $\SU(3)_C\times\SU(3)_{L}\times\SU(3)_R$, where the labels $C$, $L$, and $R$ for the $\SU(3)$ factors refer to 
\textit{color}, \textit{left} and \textit{right}, respectively. According to the trinification decomposition of the $\mathrm{E}_{6}$ adjoint representation
\begin{align}
\mathbf{78} &= (\mathbf{8},\mathbf{1},\mathbf{1}) \;\oplus\;
    (\mathbf{1},\mathbf{8},\mathbf{1}) \;\oplus\;
    (\mathbf{1},\mathbf{1},\mathbf{8}) \;\oplus\;
    (\mathbf{3},\mathbf{\bar{3}},\mathbf{\bar{3}}) \;\oplus\;
    (\mathbf{\bar{3}},\mathbf{3},\mathbf{3}), \label{eq:adjoint-into-trinification}
\end{align}
the $\mathrm{E}_6$ generators can be organized in matching order of the representations above as
\begin{align}
t_{C}^{A},\quad t_{L}^{A},\quad t_{R}^{A},\quad t^{\alpha}{}_{aa'},\quad \bar{t}_{\alpha}{}^{aa'}.\label{eq:e6-generators}
\end{align}
The indices $\alpha$, $a$ and $a'$ are the indices of the $C$, $L$ and $R$ subgroups, respectively; they run from $1$ to $3$, and are fundamental (anti-fundamental) if they are upper (lower). We also use a single label for the adjoint index $A$ running from $1$ to $8$, always in the upper position, with the subscript of the generator indicating which $\SU(3)$ factor the $A$ refers to. This notation is standard and has been used before~\cite{Kephart:1981gf,Bajc:2013qra,Babu:2015psa}.

We also follow the same references in the use of tensor formalism for irreducible representations of $\mathrm{E}_{6}$. For the (anti)-fundamental representation $\mathbf{27}$ ($\overline{\mathbf{27}}$) of $\mathrm{E}_{6}$, we use an upper (lower) index $i$ that runs from $1$ to $27$. All irreducible representations of $\mathrm{E}_{6}$ with dimension smaller than $1000$ can be accessed through the following two tensor products:
\begin{align}
    \mathbf{27}\otimes \mathbf{27}&=\overline{\mathbf{27}}_S\oplus \mathbf{351}'_S \oplus \mathbf{351}_A   , \label{eq:product-27-27}\\
    \mathbf{27}\otimes \overline{\mathbf{27}}&=\mathbf{1} \oplus \mathbf{78} \oplus \mathbf{650}, \label{eq:product-27-27bar}
\end{align}
where the subscripts $S$ and $A$ indicate the symmetric and anti-symmetric parts of the product, respectively.

\begin{table}[htb]
    \begin{center}
    \caption{List of non-trivial irreducible representations (up to conjugation) of $\mathrm{E}_{6}$ with $\mathrm{dim}<1000$. They are written as tensors with a generic label $X$ satisfying certain $\mathrm{E}_{6}$-invariant conditions. $D_{2}$ denotes the Dynkin index of the representation. \label{table:E6-irreps}}
    \begin{tabular}{lllrc}
        \toprule
        irrep & indices & conditions & $D_{2}$&self-conjugate?\\ 
        \midrule
        $\phantom{0}\mathbf{27}$&
            $X^{i}$&
            /&
            $3$&
            no\\
        $\phantom{0}\mathbf{78}$&
            $X^{i}{}_{j}$&
            $X^{i}{}_{j}=X^{K} (t^{K})^{i}{}_{j}$&
            $12$&
            yes\\
        $\mathbf{351}$&
            $X^{ij}$&
            $X^{ij}=-X^{ji}$&
            $75$&
            no\\
        $\mathbf{351}'$&
            $X^{ij}$&
            $X^{ij}=X^{ji}$, $d_{ijk}\,X^{jk}=0$&
            $84$&
            no\\
        $\mathbf{650}$&
            $X^{i}{}_{j}$&
            $\mathbf{X}=\mathbf{X}^\dagger$, $X^{i}{}_{i}=0$, $X^{i}{}_{j}\,(t^{K})^{j}{}_{i}=0$&
            $150$&
            yes\\
        \bottomrule
    \end{tabular}
    \end{center}
\end{table}

Given these decompositions, and making use of the $\mathrm{E}_{6}$ completely-symmetric invariant tensors $d_{ijk}$ and $d^{ijk}$, the irreducible representations in question can be formally written as tensors with up to $2$ indices, as shown in Table~\ref{table:E6-irreps}. The $\mathrm{E}_{6}$ generators there have been generically labeled as $t^{K}$, where the adjoint index $K$ runs from $1$ to $78$. 

Some further technical details, e.g.~the commutation relations of $\mathrm{E}_{6}$ generators and the definition of the invariant $d$-tensor, are provided in Appendix~\ref{appendix:E6}.

\subsection{Maximal Subgroups of $\mathrm{E}_{6}$ \label{sec:maximal-subgroups}}

Since $\mathrm{E}_{6}$ is a relatively large group, the possible breaking chains to the SM group $G_{321}$ become very elaborate, much more so than e.g.~in $\SO(10)$~\cite{Chang:1984qr,Deshpande:1992au,Deshpande:1992em}. In fact, since it is a subgroup of $\mathrm{E}_{6}$, all $\SO(10)$ breaking chains become subchains of the larger $\mathrm{E}_{6}$ breaking patterns. 

Our primary interest in this paper are the new breaking chains to the SM not available in $\SO(10)$ GUT, in particular breaking through $G_{333}$ and $G_{62}$. Both are examples of maximal subgroups of $\mathrm{E}_{6}$, and we shall therefore limit the investigation to such cases --- a limitation that we shall briefly revisit at the end of this section. The scope of 
maximal $\mathrm{E}_{6}$-subgroups with potential phenomenological relevance is then rounded out by $\SO(10)\times\UU$, which is covered in this work as well.

The list of maximal subgroups of $\mathrm{E}_{6}$ is provided in Table~\ref{table:E6-maximal-subgroups} (based on \cite{Feger:2019tvk,Yamatsu:2015npn}). For each subgroup in the table, we list whether it requires a regular (R) or special (S) embedding, the decomposition of the adjoint representation $\mathbf{78}$ of $\mathrm{E}_{6}$ into this subgroups' irreducible representations, the number of subgroup singlets found when decomposing the $\mathbf{650}$ of $\mathrm{E}_{6}$, and whether $G$ contains the SM group $G_{321}$ as a subgroup. In the decomposition of the $\mathbf{78}$, the parts which belong to the adjoint of the subgroup $G$ are underlined; in a spontaneous symmetry breaking scenario to that subgroup the gauge bosons that are not underlined thus become massive.

\begin{table}[htb]
    \begin{center}
    \caption{List of maximal subgroups $G$ of $\mathrm{E}_{6}$, along with embedding type (regular/special), decomposition of the $\mathbf{E}_{6}$ adjoint into irreducible representations of $G$ (with the adjoint representation(s) of $G$ underlined), and number $\#$ of $G$-singlets in the $\mathbf{650}$. The last column indicates whether the SM group is a subgroup of $G$. We consider the first $5$ cases to be the ``relevant'' subgroups. \label{table:E6-maximal-subgroups}}
    \begin{tabular}{l@{\ }l@{\quad}llrr}
    \toprule
   subgroup $G$&&type&decomposition of $\mathbf{78}$&$\#_{\mathbf{1}_G\in\mathbf{650}}$& $G_{321}\stackrel{?}{\subset }G$\\
    \midrule
    $\SU(3)\times\SU(3)\times\SU(3)$& $\equiv G_{333}$
        &R&
        $\EMPH{(\mathbf{8},{\mathbf{1},\mathbf{1}})}\oplus \EMPH{(\mathbf{1},{\mathbf{8},\mathbf{1}})}\oplus \EMPH{(\mathbf{1},{\mathbf{1},\mathbf{8}})}\oplus$
        &2&yes\\
    &&&
    $\oplus (\mathbf{3},{\mathbf{\bar{3}},\mathbf{\bar{3}}})\oplus (\mathbf{\bar{3}},{\mathbf{3},\mathbf{3}})$&&\\
    $\SU(6)\times\SU(2)$ & $\equiv G_{62}$
        &R&
        $\EMPH{(\mathbf{35},\mathbf{1})} \oplus 
        \EMPH{(\mathbf{1},\mathbf{3})} \oplus (\mathbf{20},\mathbf{2})$
        &1&yes\\
    $\SO(10)\times \mathrm{U}(1)$ & $\equiv G_{10,1}$
        &R&
        $\EMPH{(\mathbf{45},0)}\oplus \EMPH{(\mathbf{1},0)} \oplus
        (\mathbf{16},+3)\oplus (\mathbf{\overline{16}},-3)$
        &1&yes\\
    $\mathrm{F}_{4}$&
        &S&
        $\EMPH{\mathbf{56}}\oplus\mathbf{26}$
        &1&no\\
    $\SU(3)\times\mathrm{G}_{2}$&
        &S&
        $\EMPH{(\mathbf{8},\mathbf{1})}\oplus \EMPH{(\mathbf{1},\mathbf{14})} \oplus (\mathbf{8},\mathbf{7})$
        &1&no\\
    $\mathrm{G}_{2}$&
        &S&
        $\EMPH{\mathbf{14}}\oplus\mathbf{64}$
        &0&no\\
    $\SU(3)$&
        &S&
        $\EMPH{\mathbf{8}}\oplus\mathbf{35}\oplus\mathbf{\overline{35}}$
        &0&no\\
    $\mathrm{Sp}(8)$&
        &S&
        $\EMPH{\mathbf{36}}\oplus\mathbf{42}$
        &0&no\\
    \bottomrule
    \end{tabular}
    \end{center}
\end{table}

We observe from Table~\ref{table:E6-maximal-subgroups} that the SM group can be found in $G$ for the first three cases, and hence these cases are of phenomenological relevance. Incidentally, the first three cases are also exactly those with a regular embedding of $G$ into $\mathrm{E}_{6}$, i.e.~the embedding is rank-preserving.

The cases that may be obtained as the residual symmetry after first stage breaking by the vacuum expectation value (VEV) of the representation $\mathbf{650}$, however, are those which contain at least one singlet in this representation. Table~\ref{table:E6-maximal-subgroups} shows these are the regular cases, as well as the special embedding cases of $\mathrm{F}_{4}$ and $\SU(3)\times\mathrm{G}_{2}$, which involve the exceptional groups $\mathrm{F}_{4}$ and $\mathrm{G}_{2}$. It is these maximal cases with singlets in the $\mathbf{650}$ that we focus on in our vacuum analysis, and we shall refer to them as the ``relevant'' subgroups.

The ``relevant'' subgroups $G$ can be specified concretely by listing their generators as a linear combinations of those from Eq.~\eqref{eq:e6-generators}. The regular cases can be defined by subsets of $\mathrm{E}_6$ generators
\begin{align}
    \SU(3)^3:&\quad
		\{ t^{A}_{C},\ t^{A}_{L},\ t^{A}_{R} \}, \label{eq:define-rsubgroups-su3su3su3}\\
    \SU(6)\times\SU(2):&\quad
		\{ t^{A}_{C},\ t^{A}_{L},\ t^{1,2,3,8}_{R},\ t^{\alpha}{}_{a3},\ \bar{t}_{\alpha}{}^{a3}\},\label{eq:define-rsubgroups-su6su2}\\
	\SO(10)\times\UU:&\quad
		\{ t^{A}_{C}, t^{1,2,3,8}_{L}, t^{1,2,3,8}_{R}, 
			t^{\alpha}{}_{bb'}, \bar{t}_{\alpha}{}^{bb'};\ 
		(bb')=(11),(12),(21),(22),(33)\}, \label{eq:define-rsubgroups-so10}
\end{align}
while a more elaborate construction is necessary for those with a special embedding, see Appendix~\ref{appendix:special-embeddings} for technical details. In Eqs.~\eqref{eq:define-rsubgroups-su3su3su3}--\eqref{eq:define-rsubgroups-so10} the index $A$ runs unrestricted from $1$ to $8$ and indices $a$ and $a'$ from $1$ to $3$, while specific index values are listed with commas in cases when only some generators of that type are present, e.g.~$t_{L}^{1,2,3,8}$. Along with the already defined abbreviations $G_{333}$ and $G_{62}$, we shall sometimes shorten $\SO(10)\times\mathrm{U}(1)$ to $G_{10,1}$, as indicated in Table~\ref{table:E6-maximal-subgroups}.

We finish this section on subgroups with a few noteworthy remarks:
\begin{itemize}
\item
    The subgroup specifications of Eqs.~\eqref{eq:define-rsubgroups-su3su3su3}--\eqref{eq:define-rsubgroups-so10} are only one possible embedding of each into $\mathrm{E}_{6}$. Given a subalgebra embedding $\mathfrak{g}\hookrightarrow \mathfrak{e}_{6}$, we can obtain an equivalent embedding by conjugation, i.e.~$A \mathfrak{g} A^{-1}$ also constitutes an embedding of $\mathfrak{g}$ if we conjugate with any group element $A\in\mathrm{E}_{6}$. This shows there are always an infinite number of ways to embed $G$ into $\mathrm{E}_{6}$. Each case of $G$ has, however, only one embedding (of its Lie algebra) into $\mathrm{E}_{6}$ up to conjugacy, cf.~e.g.~lists of maximal algebras in~\cite{Feger:2019tvk}. The choice of one representative for each case suffices for our vacuum analysis. 
\item
    A related question concerns the embedding of the SM into $\mathrm{E}_{6}$. Given a fixed embedding of SM, conjugating the embedding of $G$ by an element $A\in\mathrm{E}_{6}$ may or may not change the SM embedding. From this point of view, although all embeddings of $G$ into $\mathrm{E}_{6}$ are equivalent, they may differ by how the SM group is embedded into $G$ (for example, how the hypercharge is embedded).    
    The classification of how SM embeds into the regular subgroups $G$ will not be relevant for us in this paper, since it has no bearing on the properties of the minima in the first stage of symmetry breaking, and as such this question is beyond the scope of this study, but will instead be presented in a follow-up.
\item 
    In the case of breaking chains proceeding through trinification $G_{333}$, there is an additional discrete symmetry that plays a role, as we shall see explicitly in Section~\ref{sec:potential}. We denote this symmetry by $D_{3}$. It essentially permutes the three group factors of $G_{333}$ (and is hence isomorphic to the permutation group $S_{3}$), and is generated by three parities that exchange pairs of factors, dubbed as LR, CL and CR parity. One of these parities from $D_{3}$ survives the breaking to $G_{333}$ via the $\mathbf{650}$. Note that this phenomenon is analogous to $D$-parity in the $\SO(10)$ context~\cite{Kibble:1982dd,Chang:1983fu,Chang:1984uy}, and in fact the $\mathrm{E}_{6}$ LR parity is equivalent to the $\SO(10)$ $D$-parity. We relegate the technical details and further discussion of the discrete group $D_{3}$ to Appendix~\ref{app:discrete-symmetries}.
\item
    Finally, we elaborate upon our limitation of analyzing vacua only for the ``relevant'' subgroups of Table~\ref{table:E6-maximal-subgroups}. If one uses a single irreducible representation for spontaneous symmetry breaking, as we do in the first stage breaking, Michel's conjecture states that the resulting symmetry is a maximal little group of the representation, see~\cite{Michel:1971th,Slansky:1981yr} and references therein. This suggests that large subgroups (such as maximal subgroups) are indeed of greatest interest in the analysis. Namely, vacua breaking to subgroups of the ``relevant'' ones would be in violation of Michel. The cases allowed by Michel that we do not analyze are the maximal subgroups of the last three cases in Table~\ref{table:E6-maximal-subgroups}: all of these indeed contain singlet(s) in the $\mathbf{650}$, and they are maximal little groups if they are not subgroups of the ``relevant'' cases. \\[6pt]
    Although a complete classification of all breaking solutions of a single $\mathbf{650}$ of $\mathrm{E}_{6}$ is beyond the scope of this paper, the arguments above suggest the ``relevant'' cases may indeed be the only pertinent ones, at least phenomenologically.  
\end{itemize}

\section{Analysis of scalar potential with one $\mathbf{650}$ of $\mathrm{E}_{6}$ \label{sec:potential}}

\subsection{The scalar potential \label{sec:potential-preliminaries}}

As seen in Eq.~\eqref{eq:product-27-27bar}, the representation $\mathbf{650}$ is found in the tensor product $\mathbf{27}\otimes \mathbf{\overline{27}}$. It can thus be implemented as a $27\times 27$ matrix $\mathbf{X}$, written as $X^{i}{}_{j}$ in tensor notation, with some additional constraints placed on it, cf.~Table~\ref{table:E6-irreps}. 

The most general renormalizable potential with the representation $\mathbf{650}\equiv\mathbf{X}$ can be written as 
\begin{align}
\begin{split}
V_{650}(\mathbf{X})&=\phantom{+}M^2\cdot \mathrm{Tr}(\mathbf{X}^2)\\
&\quad +\; m_1\cdot \mathrm{Tr}(\mathbf{X}^3) +m_2\cdot X^{i}{}_{l}\,X^{j}{}_{m}\,X^{k}{}_{n}\; d^{lmn}d_{ijk}\\
&\quad +\; \lambda_1\cdot (\mathrm{Tr}(\mathbf{X}^2))^2
 \;+\; \lambda_2\cdot \mathrm{Tr}(\mathbf{X}^4)
 \;+\; \lambda_3\cdot (\mathbf{X}^2)^{k}{}_{i}\;(\mathbf{X}^2)^{l}{}_{j}\;D^{ij}{}_{kl}\\
&\quad +\;\lambda_4\cdot X^{i}{}_{i'}\,X^{j}{}_{j'}\,X^{k}{}_{k'}\,X^{l}{}_{l'}\;D^{i'j'}{}_{kl}\;D^{k'l'}{}_{ij}
 \;+\;\lambda_5\cdot X^{i}{}_{l}\,X^{j}{}_{m}\,(\mathbf{X}^2)^{k}{}_{n}\; d^{lmn}d_{ijk},\\
\end{split}\label{eq:explicit-potential}
\end{align}
where the bold-typed $\mathbf{X}$ indicates its matrix form, $X^{i}{}_{j}$ the components of this matrix, while $d_{ijk}$ is the primitive completely-symmetric invariant tensor in $\mathrm{E}_{6}$ and $D^{ij}{}_{kl}$ is a composite defined by
\begin{align}
	D^{ij}{}_{kl}&:= d^{ijm}\,d_{klm}.
\end{align}
A few more technical details for the $d$-tensors are given in Appendix~\ref{appendix:E6}. All indices involved in the above expression run from $1$ to $27$, and all parameters $M^{2}$, $m_{k}$ and $\lambda_{l}$ are real.

Notice that there are $2$ cubic and $5$ quartic invariants in Eq.~\eqref{eq:explicit-potential}. This result is non-trivial, since there can be many different ways in which the $\mathrm{E}_{6}$ indices between the tensor $X^{i}{}_{j}$ and the invariant tensors $d$ can be contracted, and furthermore the different contractions may be related to one another. This difficulty can be solved if one knows the number of independent invariants in advance, for which we used the \textit{symmetrised tensor power} function of the computer algebra program LiE~\cite{Leeuwen1992LiE}. Having access to the explicit form of the tensor $d$ then allows for constructing that many same-order invariants and checking their linear independence explicitly.

To analyze the stationarity conditions of this potential, we need to consider only the singlet directions in $\mathbf{X}$ with respect to the ``relevant'' subgroups, as discussed in Section~\ref{sec:maximal-subgroups}. We thus digress to specifying these directions below.

For the regular cases in Table~\ref{table:E6-maximal-subgroups}, a singlet of 
$G$ will also be a singlet under the SM group $G_{321}$. The singlet directions of regular $G$ are thus most conveniently viewed as directions in the sub-space of SM-singlets. The representation $\mathbf{650}$ contains $11$ real singlets of the SM group, which we label by
\begin{align}
    \{\, s,\quad a,\quad z,\quad x_1,\quad x_2,\quad X_{3},\quad y_1,\quad y_2,\quad Y_{3}\,\},\label{eq:650-SM-singlets}
\end{align}
where small letters are used for real VEVs and capital letters for complex valued VEVs containing $2$ real degrees of freedom. They are partially specified by listing their $G_{333}$ origin in Table~\ref{table:650-singlet-content}, while their explicit definition in terms of entries $X^{i}{}_{j}$ is given in Appendix~\ref{appendix:states-in-650}. 

As seen from the $G_{333}$ decomposition in Table~\ref{table:650-singlet-content}, the representation $\mathbf{650}$ has two $G_{333}$ singlets labeled by $s$ and $a$, standing for \textit{symmetric} and \textit{anti-symmetric} under left-right (LR) parity. This parity exchanges the $\SU(3)_L$ and $\SU(3)_R$ subgroups, see Appendix~\ref{app:discrete-symmetries} for further elaboration. In the decomposition to $G_{10,1}$ and $G_{62}$, however, there is only one singlet state in the $\mathbf{650}$. We use a generic label $\tilde{s}$ for this state, and it can be specified in each case as a linear combination of the SM-singlets listed in Eq.~\eqref{eq:650-SM-singlets}. Explicitly, for the subgroups defined in Eqs.~\eqref{eq:define-rsubgroups-su6su2}--\eqref{eq:define-rsubgroups-so10}, the computed directions are given in Table~\ref{table:650-subgroup-VEV-alignments}. 

The $G$-singlet directions for the special embedding cases of $\mathrm{F}_{4}$ and $\SU(3)\times\mathrm{G}_{2}$ do not live in the SM-singlet subspace, since these groups don't contain the SM group. We still label the VEV-direction in the $\mathbf{650}$ that retains these symmetries by $\tilde{s}$, but we relegate their explicit definition to Appendix~\ref{appendix:states-in-650}. 

Note that the VEVs in the $G$-singlet direction(s) are always normalized such 
that the following hold:
\begin{align}
     \langle \mathrm{Tr}(\mathbf{X}^2)\rangle &= \tfrac{1}{2}\,(s^{2}+a^{2}), &
    \langle \mathrm{Tr}(\mathbf{X}^2)\rangle&=\tfrac{1}{2}\,\tilde{s}^2.
\end{align}
The former applies to $G_{333}$, and the latter to all other cases with only one $G$-singlet in the $\mathbf{650}$.

\begin{table}[htb]
	\centering
	\caption{The $G_{333}$-locations of 11 SM real singlets from the $\mathbf{650}$ of $\mathrm{E}_{6}$. The two states $s$ and $a$ have a well-defined LR parity. Explicit definitions of states can be found in Appendix~\ref{appendix:states-in-650}. 	\label{table:650-singlet-content}}
	\begin{tabular}{lcc}
		\toprule
		$G_{333}$ irrep&LR parity&SM singlets inside\\
		\midrule
		$(\mathbf{1},\mathbf{1},\mathbf{1})$&$+1$&$s$\\
		$(\mathbf{1},\mathbf{1},\mathbf{1})$&$-1$&$a$\\
		$(\mathbf{1},\mathbf{1},\mathbf{8})$&&$x_1$, $x_2$, $\mathrm{Re}(X_3)$, $\mathrm{Im}(X_3)$\\
		$(\mathbf{1},\mathbf{8},\mathbf{1})$&&$z$\\
		$(\mathbf{1},\mathbf{8},\mathbf{8})$&&$y_1$, $y_2$, $\mathrm{Re}(Y_3)$, $\mathrm{Im}(Y_3)$\\
		\bottomrule
	\end{tabular}
\end{table}

\begin{table}[htb]
	\centering
	\caption{The VEV alignment of $\tilde{s}$ in the $\mathbf{650}$ which breaks $E_6$ to the regular subgroups $G$ defined in Eqs.~\eqref{eq:define-rsubgroups-su3su3su3}--\eqref{eq:define-rsubgroups-so10}. The $G_{333}$ case has $2$ possible independent directions $s$ and $a$ for the VEV. The special cases $\mathrm{F}_{4}$ and $\SU(3)\times\mathrm{G}_{2}$ are considered separately in Appendix~\ref{appendix:states-in-650}. \label{table:650-subgroup-VEV-alignments}}
	\begin{tabular}{ll}
		\toprule
		subgroup $G$&VEV alignment\\
		\midrule
        $G_{333}$&$s$, $a$ \\
        $G_{62}$&$(s,a,x_1,x_{2})=
			\tfrac{1}{2\sqrt{30}}\,(2\sqrt{3},6,-3\sqrt{2},-3\sqrt{6})\;\tilde{s}$ \\
		$G_{10,1}$&$(s,z,x_1,x_2,y_1,y_2)=\tfrac{1}{\sqrt{80}}\,(2\sqrt{2},2\sqrt{3},\sqrt{3},3,-2\sqrt{3},6)\;\tilde{s}$\\
\bottomrule
\end{tabular}
\end{table}

Returning to the analysis of the scalar potential $V_{650}$ in Eq.~\eqref{eq:explicit-potential}, the stationarity conditions can be 
expressed only in terms of the defined singlet directions for each of the relevant cases of Table~\ref{table:E6-maximal-subgroups}. The $G$-non-singlets, on the other hand,  do not need to be considered since they have a vanishing VEV --- we refer to the potential with all non-singlet zero-VEVs inserted as the \textit{restricted potential}. It turns out all cases lead to a similar expression for the restricted potential that takes the general form
\begin{align}
	V(s,a)&=\tfrac{1}{2}M^2\;(s^2+a^2)-\tfrac{1}{3}m\;s(s^2-3a^2)+\tfrac{1}{4}\lambda\;(s^2+a^2)^2,\label{eq:singlet-potential-1}
\end{align}
where the parameter $M^{2}$ is the same as in $V_{650}$, while $m$ and $\lambda$ are determined from the original parameters in Eq.~\eqref{eq:explicit-potential} via
\begin{align}
	\label{eq:parameters}
	m&:=\sum_{k=1}^{2}\alpha_k\, m_k, & \lambda&:= \sum_{l=1}^{5}\beta_l\,\lambda_l,
\end{align}
where the numbers $\alpha_{k}$ and $\beta_{l}$ are subgroup dependent and are given in Table~\ref{table:vacua-alfa-beta}. For all cases other than trinification, when only one singlet is present, one obtains the restricted pontential by setting $s=\tilde s$ and $a=0$. 

The last column of Table~\ref{table:vacua-alfa-beta} also gives explicit expressions for the masses of those gauge bosons which correspond to broken symmetry generators, i.e.~those representations of $G$ not underlined in the decomposition of the $\mathbf{78}$ in Table~\ref{table:E6-maximal-subgroups}. These results come from the computation of all $78$ gauge boson masses in each case. They  thus offer an independent check that we correctly identified the singlets of $G$ in the representation $\mathbf{650}$, since we obtain the correct counting of degeneracies in gauge boson masses. 

We analyze the restricted potential of Eq.~\eqref{eq:singlet-potential-1} and obtain minima candidates for the potential in Eq.~\eqref{eq:explicit-potential} in the next subsection.

\begin{table}[htb]
	\centering
	\caption{Coefficients $\alpha_k$ and $\beta_l$ needed to compute the parameters of the potential in Eq.~\eqref{eq:singlet-potential-1} using 
Eq.~\eqref{eq:parameters}. The last column gives masses acquired by the gauge bosons once the VEVs are engaged.
	\label{table:vacua-alfa-beta}}
	\begin{tabular}{llll}
		\toprule
		subgroups&$\alpha_k$&$\beta_l$&value of $M^{2}_{GUT}$\\
		\midrule
        $G_{333}$&$\tfrac{1}{4\sqrt{3}}(1,10)$&$\tfrac{1}{18}(18,1,7,34,4)$&$M^{2}_{(\mathbf{3},\mathbf{\bar{3}},\mathbf{\bar{3}})}=\tfrac{1}{3}\,g^2(s^2+a^2)$\\[6pt]
        $G_{62}$&$\tfrac{1}{4\sqrt{30}}(1,-44)$&$\tfrac{1}			{180}(180,7,67, 652, -8)$&$M^{2}_{(\mathbf{20},\mathbf{2})}=\tfrac{9}{20}\, g^2 \tilde{s}^2$\\[6pt]
		$G_{10,1}$&$\tfrac{1}{8\sqrt{30}}(29,20)$&$\tfrac{1}{720}(720,397,97,520,40)$&$M^{2}_{\mathbf{16}(-3)}=\tfrac{9}{16}\, g^2 \tilde{s}^2$\\[6pt]
		$\mathrm{F_{4}}$&$\tfrac{1}{\sqrt{78}}(25,34)$&$\tfrac{1}{234}(234,217,295,394,292)$& $M^{2}_{\mathbf{26}}=\tfrac{9}{13}\,g^{2} \Tilde{s}^{2}$\\[6pt]
		$\SU(3)\times \mathrm{G}_{2}$&$\tfrac{1}{\sqrt{42}}(5,23)$&$\tfrac{1}{126}(126,13,55,181,43)$& $M^{2}_{(\mathbf{8},\mathbf{7})}=\tfrac{9}{28}\,g^{2} \Tilde{s}^{2}$\\
\bottomrule
\end{tabular}
\end{table}

\subsection{Analysis of the vacua \label{sec:analysis-of-vacua}}

\subsubsection{Solution to stationarity conditions \label{sec:vacuum-solutions}}

We start by analyzing the most general form of the restricted potential in Eq.~\eqref{eq:singlet-potential-1}. Since the VEVs $s$ and $a$ are real, the potential is a polynomial function mapping $\mathbb{R}^{2}\to\mathbb{R}$. For the potential to be bounded from below, we demand $\lambda>0$. 

First, we notice a three-fold symmetry in the restricted potential. Suppose that we perform a rotation in the $(s,a)$ plane by an angle $\varphi$:
\begin{align}
	\begin{pmatrix} s\\ a\\ \end{pmatrix}
		&\mapsto R(\varphi)
	\begin{pmatrix} s\\ a\\ \end{pmatrix},
 &
	R(\varphi)&:=\begin{pmatrix}
		\cos\varphi&-\sin\varphi\\
		\sin\varphi&\phantom{-}\cos\varphi\\
		\end{pmatrix}. \label{eq:R-rotation}
\end{align}
Inserting this transformation into the restricted potential, we get
\begin{align}
	V_{650}(s,a)&\mapsto \tfrac{1}{2}M^2\;(s^2+a^2)-\tfrac{1}{3}m\;\left(s(s^2-3a^2)
\cos(3\varphi)+ a(a^2-3s^2)\sin(3\varphi)\right)+\tfrac{1}{4}\lambda\;(s^2+a^2)^2.
\end{align}
Angles that satisfy $3\varphi=2\pi k$ for integer $k$ clearly yield back the original restricted potential of Eq.~\eqref{eq:singlet-potential-1}. The potential $V$ thus exhibits a $\mathbb{Z}_3$ symmetry in the $s$-$a$ plane, generated by $R(2\pi/3)$. Furthermore, the flip $a\mapsto -a$ generates a $\mathbb{Z}_{2}$ parity transformation, which also leaves the restricted potential invariant. All together these transformations form 
an $S_3$ permutation symmetry of the restricted potential. 

The full two-singlet expression of the restricted potential is relevant only for the $G_{333}$ trinification case, and in this context the $S_{3}$ transformations correspond to the discrete group $D_{3}$ from Section~\ref{sec:maximal-subgroups}. To reiterate, it is generated by the $\mathbb{Z}^{LR}_{2}$, $\mathbb{Z}^{CL}_{2}$ and $\mathbb{Z}^{CR}_{2}$ parities defined in Appendix~\ref{app:discrete-symmetries}. The flip $a\mapsto -a$ directly corresponds to $\mathbb{Z}^{LR}_{2}$ parity, in accordance with Table~\ref{table:650-singlet-content}.

We now proceed by analyzing the restricted potential directly: the stationarity conditions 
\begin{align}
\tfrac{\partial}{\partial s}V=\tfrac{\partial}{\partial a}V&=0
\end{align}
\noindent
yield $7$ formal solutions $(s_i,a_i)$, with $i=0,\ldots 6$ as follows:
\begin{align}
	(s_0,a_0)&=(0,0), \label{eq:sa-solution-0}\\
	(s_{1,2},a_{1,2})&=(\tfrac{m\pm\sqrt{m^2-4\lambda\,M^2}}{2\lambda},0), 
	\label{eq:sa-solution-1}
\end{align}
while the remaining $4$ solutions are the $(s_j,a_j)$ solutions for $j=1,2$ rotated by either $R(2\pi/3)$ or $R(4\pi/3)$ due to the $\mathbb{Z}_3$ symmetry of the restricted potential. When the formal expressions for the solution yield real values, i.e.~ when the discriminant $\mathcal{D}$ is positive
\begin{align}
\mathcal{D}&:=m^2-4\lambda\,M^2 >0,
\end{align}
non-trivial minima can exist. 

Intuitively, demanding $\lambda>0$ and destabilizing the trivial solution by $M^2<0$, the discriminant is positive for all $m\in\mathbb{R}$. Since at least one of the points must be a local minimum in such a case, combining $\mathbb{Z}_3$ with the Poincare-Hopf index theorem results in having $3$ minima and $3$ saddle points. 

The pattern of minima is confirmed in the general case by the Hessian matrix of $V$ with the inserted $(s_{1,2},a_{1,2})$ solutions:
	\begin{align}
	\label{eq:ddV}
	\partial^{2}V\Big|_{(s,a)=(s_{1,2},a_{1,2})}&=\frac{m\pm\sqrt{\mathcal{D}}}{2\lambda}\;\mathrm{diag}			\left(\pm\sqrt{\mathcal{D}},3m\right).
\end{align}
For $m>0$, together with the non-trivial minima and boundedness conditions $\lambda,\mathcal{D}>0$, we therefore see that $(s_i,a_i)$ is a (global) minimum of the restricted potential $V$ for $i=1$ (the point with $s>0$), and a saddle point for $i=2$. For $m<0$, the role of minima and saddle points are exchanged.\footnote{Note that the sign of $\mathbf{650}$ can be redefined it necessary, so that one can take $m>0$ without loss of generality. The parameter $m$ is, however, a derived parameter dependent on $G$, so the redefinition is possible only for an analysis limited to vacua of a single $G$, while $m>0$ may not be achieved for all $G$ simultaneously.} Also, if $M^{2}=0$, the saddle points merge with $P_{0}$ for any sign of $m$.

\begin{figure}[htb]
	\centering
	\includegraphics[width=10cm]{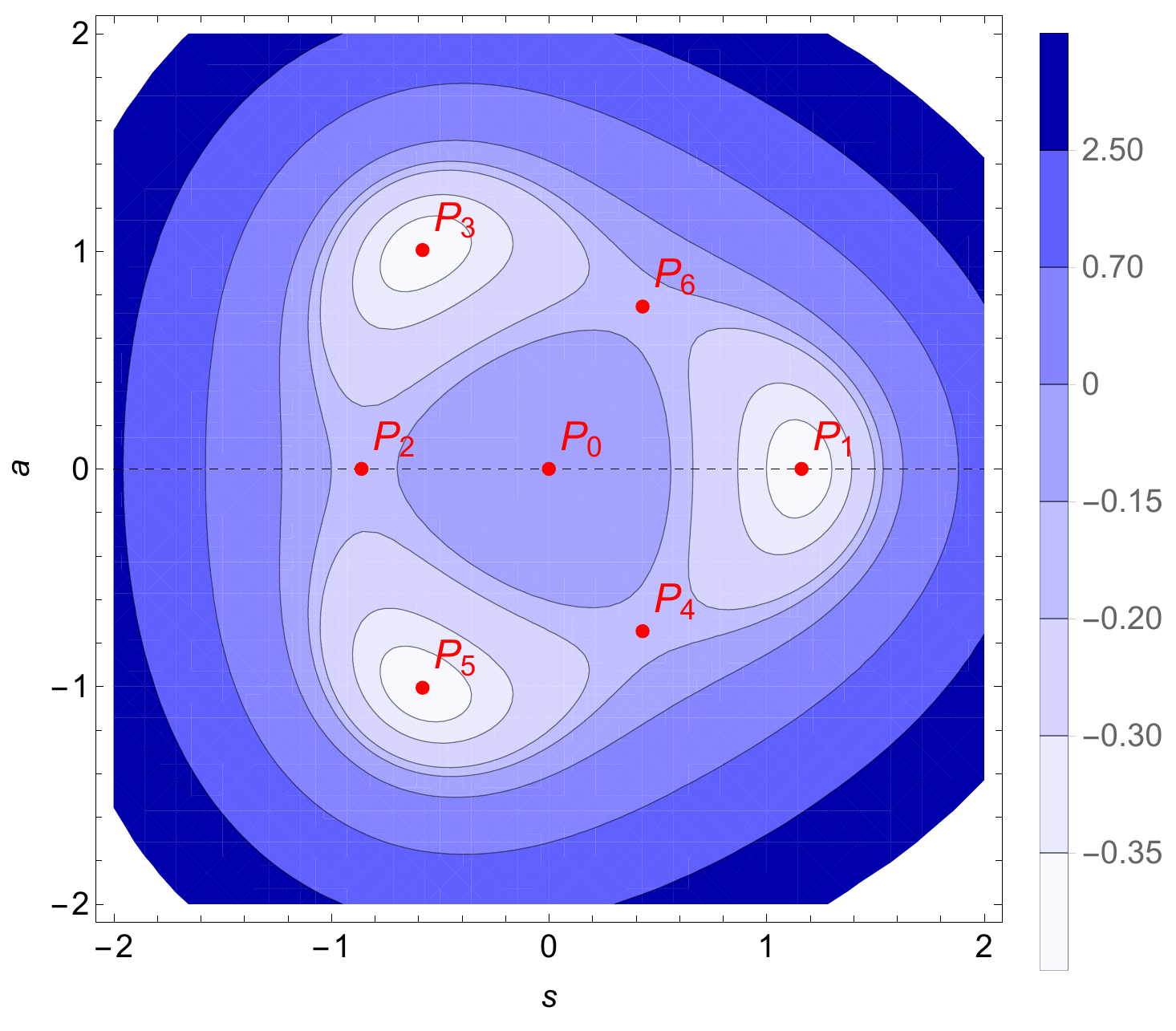}
	\caption{The contour plot in the $(s,a)$-plane for the restricted potential $V$ of Eq.~\eqref{eq:singlet-potential-1} taking $M^{2}=-1$, $m=0.3$ and $\lambda=1$ (using arbitrary mass units). The stationary points $P_{i}=(s_{i},a_{i})$ are shown in red.\label{fig:restricted-potential}}
\end{figure}

The potential $V$ is visualized in the contour plot of the restricted potential in Figure~\ref{fig:restricted-potential}. Specific parameter values $(M^{2},m,\lambda)=(-1,0.3,1)$ were chosen. The stationary points are indicated as red dots and denoted by $P_{i}=(s_{i},a_{i})$ for $i=0,\ldots,6$. The plot clearly exhibits the three-fold $\mathbb{Z}_{3}$ rotational symmetry, and it is symmetric under the $a\mapsto -a$ reflection. The point $P_{0}$ is a local maximum, $P_{1,3,5}$ are global minima, and $P_{2,4,6}$ are saddle points. 

Suppose we define two new rotated sets of coordinates beside the pair $(s,a)$:
\begin{align}
	(s',a')^{T}&:=R(2\pi/3)\,(s,a)^{T}, &
	(s'',a'')^{T}&:=R(4\pi/3)\,(s,a)^{T}.
\end{align}
We see that the points $P_2$, $P_0$ and $P_{1}$ lie on the $a=0$ line. Similarly, the points $P_{4,0,3}$ lie on the $a'=0$ line and $P_{6,0,5}$ lie on the $a''=0$ line due to rotational symmetry. The implication for trinification is that
each of the $3$ minima preserves a different $\mathbb{Z}_2$ parity: $\{P_{1},P_{3},P_{5}\}$ preserve the $\{$LR,CL,CR$\}$ parities corresponding to $\{a=0,a'=0,a''=0\}$, respectively. In particular, any trinification vacuum preserves a particular $\mathbb{Z}_{2}$ parity remnant of the discrete symmetry $D_{3}$ of the restricted potential.

For all cases of subgroups $G$ other than trinification, when there is only one singlet $\tilde{s}$, the stationarity condition is a cubic equation of $\tilde{s}$ and therefore has $3$ formal solutions. These consist of setting $s=\tilde s$ and $a=0$ in Eqs.~\eqref{eq:sa-solution-0}--\eqref{eq:sa-solution-1}, with the second derivative at $\tilde{s}_{1,2}$ equal to the upper-left entry in the matrix of Eq.~\eqref{eq:ddV}. 
\begin{itemize}
\item 
    In the regime $M^{2}<0$ and $\lambda>0$, we always have $\mathcal{D}> m^{2}$, and thus the solution $\tilde{s}_{0}=0$ is a local maximum and $\tilde{s}_{1,2}$ are two local minima. For $m>0$, the $+$ solution $\tilde{s}_{1}$ is the global minimum of $V$. This can be visualized in Figure~\ref{fig:restricted-potential} if we restrict to the dashed line $a=0$: $P_{1}$ and $P_{2}$ are local minima along this line, with $P_{1}$ being also the global minimum. 
\item 
    In the alternative regime  $M^{2},\lambda>0$, the trivial solution is a local minimum and non-trivial solutions exist only for $\mathcal{D}>0$; in such a case we have $|m|>\sqrt{\mathcal{D}}$, and so for $m>0$ the solution $\tilde{s}_{1}$ is a local minimum, while $\tilde{s}_{2}$ is a local maximum.
\item For $M^{2}=0$, the trivial minimum and local maximum from the $M^{2}>0$ case merge into a saddle point, and one local minimum remains ($s_{1}$ for $m>0$). 
\end{itemize}

To summarize, for all symmetries $G$ there is at least one corresponding (local) minimum of the restricted potential when $\lambda,\mathcal{D}>0$. For $G_{333}$ there are $3$ degenerate minima, and for all other cases of $G$ in Table~\ref{table:vacua-alfa-beta} we have either one ($M^{2}\geq 0$) or two ($M^{2}<0$) local minima. In the latter case, their difference in depth is controlled by the parameter $m$ of the cubic term. 

We stress that these results hold for the restricted potential. In the full potential $V_{650}$ of Eq.~\eqref{eq:explicit-potential}, one also has to consider the fields in non-singlet directions. By construction, these were taken with zero VEVs and the local minima of the restricted potential $V$ are stationary points of the full potential $V_{650}$. Their ultimate fate, i.e.~whether they become local minima or saddle points of $V_{650}$, is addressed in Section~\ref{sec:scalar-spectrum}.

\subsubsection{The scalar spectrum in different vacua \label{sec:scalar-spectrum}}

Armed with the solutions for stationarity points for different cases of subgroup $G$, the obvious next step is to determine the conditions under which these are (local) minima of the potential $V$ in Eq.~\eqref{eq:explicit-potential} rather than merely saddle points. In other words, we are seeking conditions under which the potential is stabilized in all $650$ field directions. This is easily addressed by considering the masses of all the states in the $\mathbf{650}$, which can be grouped according to the residual symmetry $G$.

We provide the masses of representations in different vacua in a series of tables. 
The masses for the $G_{333}$ vacuum are listed in Table~\ref{tab:masses-vacuum-su3su3su3}, for the $G_{62}$ case in Table~\ref{tab:masses-vacuum-su6su2},
and for the $G_{10,1}$ case in Table~\ref{tab:masses-vacuum-so10u1}. 
In all tables we also specify for each representation $R$ whether the fields inside are complex or real; the reader can verify in each case that the total number of real degrees of freedom adds up to $650$. Since the $\mathrm{F}_{4}$ and $\SU(3)\times\mathrm{G}_{2}$ vacua are not of phenomenological relevance, we omit the computation of associated masses for these cases. 

The results in the tables were obtained by differentiating the full potential of Eq.~\eqref{eq:explicit-potential} with respect to two appropriately-transforming  fields $\phi_{a}$ and $\phi_{b}$ (components of the in principle complex representation $R$ of $G$) from $\mathbf{X}\sim\mathbf{650}$, and inserting the vacuum condition, i.e.~$(m^{2})^{a}{}_{b}=\partial^{a\ast} \partial_{b} V_{650}|$. In order to obtain simpler expressions, we solved the stationarity conditions for $M^{2}$ instead of the VEV $s$ (or $\tilde{s}$). Note that for a fixed parameter point $(M^{2},m_{k},\lambda_l)$, the VEVs take different values in different tables. Explicitly, $s$ in Tables~\ref{tab:masses-vacuum-su3su3su3} and $\tilde{s}$ in Tables~\ref{tab:masses-vacuum-su6su2} and \ref{tab:masses-vacuum-so10u1} take the value of the solution in Eq.~\eqref{eq:sa-solution-1} that corresponds to a local minimum of the restricted potential $V$ (identified by $\partial^{2}V$ via Eq.~\eqref{eq:ddV}), where the parameters $m$ and $\lambda$ are $G$-dependent and are determined from Eq.~\eqref{eq:parameters} and Table~\ref{table:vacua-alfa-beta}. For trinification masses in Table~\ref{tab:masses-vacuum-su3su3su3}, we chose the LR-symmetric vacuum with $a=0$ out of the three discrete possibilities.

\begin{table}[htb]
\begin{center}
\caption{Masses of irreducible representations in the $\SU(3)_{C}\times\SU(3)_{L}\times\SU(3)_{R}$ vacuum with LR parity. While $a=0$ has already been inserted,
$s$ is a local minimum of $V$ determined from Eqs.~\eqref{eq:sa-solution-1}--\eqref{eq:ddV}, where $m$ and $\lambda$ are computed from Eq.~\eqref{eq:parameters} and Table~\ref{table:vacua-alfa-beta}.
\label{tab:masses-vacuum-su3su3su3}}
\begin{tabular}{l@{\quad}c@{\quad}l}
    \toprule
    irreps $R$ of $G_{333}$&$\mathbb{R}/\mathbb{C}$&masses-square $m^{2}_{R}$\\
    \midrule
    $(\mathbf{1},\mathbf{1},\mathbf{1})_{1}$ & 
    $\mathbb{R}$ &
    $\tfrac{1}{36} s \left(4 s (18 \lambda_{1}+\lambda_{2}+7 \lambda_{3}+34 \lambda_{4}+4 \lambda_{5})-3 \sqrt{3} (m_{1}+10 m_{2})\right)$\\
    $(\mathbf{1},\mathbf{1},\mathbf{1})_{2}$ &
    $\mathbb{R}$ &
    $\tfrac{1}{4} \sqrt{3} s (m_{1}+10 m_{2}) $\\
    $(\mathbf{8},\mathbf{1},\mathbf{1})$ &
    $\mathbb{R}$ &
    $\tfrac{1}{4} s \left(\sqrt{3} m_{1}-2 \sqrt{3} m_{2}+2s(4\lambda_{4}- \lambda_{5})\right)$\\
    $(\mathbf{1},\mathbf{8},\mathbf{1})$, $(\mathbf{1},\mathbf{1},\mathbf{8})$ &
    $\mathbb{R}$ &
    $\tfrac{1}{12} s \left(18 \sqrt{3} m_{2}+s (\lambda_{2}-5\lambda_{3}-20 \lambda_{4}-5\lambda_{5})\right)$ \\
    $(\mathbf{3},\mathbf{\bar{3}},\mathbf{\bar{3}})_{1}$ &
    $\mathbb{C}$ &
    $\tfrac{1}{18} s \left(3 \sqrt{3} (m_{1}+10 m_{2})-2 s (\lambda_{3}+2 (8 \lambda_{4}+\lambda_{5}))\right)$ \\
    $(\mathbf{3},\mathbf{\bar{3}},\mathbf{\bar{3}})_{2}$ &
    $\mathbb{C}$ &
    $0$ \\
    $(\mathbf{3},\mathbf{6},\mathbf{\bar{3}})$, $(\mathbf{3},\mathbf{\bar{3}},\mathbf{6})$ &
    $\mathbb{C}$ &
    $s\left(\sqrt{3} m_{2}-\tfrac{1}{6} s (12 \lambda_{4}+\lambda_{5})\right)$ \\
    $(\mathbf{6},\mathbf{3},\mathbf{3})$ &
    $\mathbb{C}$ &
    $\tfrac{1}{12} s \left(3 \sqrt{3} (m_{1}+2 m_{2})-4 s (4 \lambda_{4}+\lambda_{5})\right)$ \\
    $(\mathbf{8},\mathbf{8},\mathbf{1})$, $(\mathbf{8},\mathbf{1},\mathbf{8})$ &
    $\mathbb{R}$ &
    $\tfrac{1}{4} s \left(\sqrt{3} (m_{1}+4 m_{2})-s (8 \lambda_{4}+\lambda_{5})\right)$ \\
    $(\mathbf{1},\mathbf{8},\mathbf{8})$ &
    $\mathbb{R}$ &
    $\tfrac{1}{12} s \left(2 s (\lambda_{2}+\lambda_{3}-8 \lambda_{4}+\lambda_{5})-3 \sqrt{3} (m_{1}-2 m_{2})\right)$ \\
    \bottomrule
\end{tabular}
\end{center}
\end{table}

\begin{table}[htb]
\begin{center}
\caption{Masses of irreducible representations in the $\SU(6)\times\SU(2)$ vacuum, where $\tilde{s}$ is a local minimum of $V$ determined from Eqs.~\eqref{eq:sa-solution-1}--\eqref{eq:ddV}, where $m$ and $\lambda$ are computed from Eq.~\eqref{eq:parameters} and Table~\ref{table:vacua-alfa-beta}. \label{tab:masses-vacuum-su6su2}}
\begin{tabular}{r@{\quad}c@{\quad}l}
    \toprule
    irrep $R$ of $G_{62}$&$\mathbb{R}/\mathbb{C}$&mass-square $m^{2}_{R}$\\
    \midrule
    $(\mathbf{1},\mathbf{1})$ &
    $\mathbb{R}$ & 
    $\tfrac{1}{360} \tilde{s}\, \left(4 \tilde{s}\, (180 \lambda_{1}+7 \lambda_{2}+67 \lambda_{3}+652 \lambda_{4}-8 \lambda_{5})-3 \sqrt{30}\, (m_{1}-44 m_{2})\right)$ \\
    $(\mathbf{35},\mathbf{1})$ &
    $\mathbb{R}$ & 
    $\tfrac{1}{120} \tilde{s} \,\left(2 \tilde{s}\, (5 \lambda_{2}-19 \lambda_{3}-208 \lambda_{4}+2 \lambda_{5})-3 \sqrt{30}\, (m_{1}+28 m_{2})\right)$ \\
    $(\mathbf{20},\mathbf{2})$ &
    $\mathbb{R}$ & 
    $0$ \\
    $(\mathbf{35},\mathbf{3})$ &
    $\mathbb{R}$ & 
    $\tfrac{1}{40} \tilde{s}\, \left(4 \tilde{s}\, (\lambda_{2}+\lambda_{3}-32 \lambda_{4}-2\lambda_{5})-3 \sqrt{30}\, (m_{1}+4 m_{2})\right)$ \\
    $(\mathbf{70},\mathbf{2})$ &
    $\mathbb{C}$ & 
    $-\tfrac{3}{40} \tilde{s}\, \left(6 \sqrt{30}\, m_{2}+\tilde{s}\, (52\lambda_{4}- \lambda_{5})\right)$ \\
    $(\mathbf{189},\mathbf{1})$ &
    $\mathbb{R}$ & 
    $\tfrac{1}{40} \tilde{s}\, \left(3 \sqrt{30}\, (m_{1}-4 m_{2})+2 \tilde{s}\, (\lambda_{2}+\lambda_{3}-64 \lambda_{4}+6 \lambda_{5})\right)$ \\
    \bottomrule
\end{tabular}
\end{center}
\end{table}

\begin{table}[htb]
\begin{center}
\caption{Masses of irreducible representations in the $\SO(10)\times\mathrm{U}(1)$ vacuum, where $\tilde{s}$ is a local minimum of $V$ determined from Eqs.~\eqref{eq:sa-solution-1}--\eqref{eq:ddV}, where $m$ and $\lambda$ are computed from Eq.~\eqref{eq:parameters} and Table~\ref{table:vacua-alfa-beta}. \label{tab:masses-vacuum-so10u1}}
\begin{tabular}{r@{\quad}c@{\quad}l}
    \toprule
    irrep $R$ of $G_{10,1}$&$\mathbb{R}/\mathbb{C}$& mass-square $m^{2}_{R}$\\
    \midrule
    $(\mathbf{1},+0)$ &
    $\mathbb{R}$&
    $\tfrac{1}{720} \tilde{s} \left(-3 \sqrt{30} (29 m_{1}+20 m_{2})+ 2 \tilde{s} (720 \lambda_{1}+397 \lambda_{2}+97 \lambda_{3}+520 \lambda_{4}+40 \lambda_{5})\right)$ \\
    $(\mathbf{10},+6)$ &
     $\mathbb{C}$&
    $\tfrac{1}{240} \tilde{s} \left(3 \sqrt{30} (4 m_{2}-5 m_{1})+ \tilde{s} (65 \lambda_{2}+185 \lambda_{3}-112 \lambda_{4}+26 \lambda_{5})\right)$ \\
    $(\mathbf{16},-3)$ &
     $\mathbb{C}$&
    $0$ \\
    $(\mathbf{45},+0)$ &
     $\mathbb{R}$&
    $\tfrac{1}{80} \tilde{s} \left(9 \sqrt{30} (m_{1}+4 m_{2}) -6 \tilde{s} (7 \lambda_{2}-5 \lambda_{3}-8 \lambda_{4}+4 \lambda_{5})\right)$ \\
    $(\mathbf{54},+0)$ &
     $\mathbb{R}$&
    $\tfrac{1}{240} \tilde{s} \left(21 \sqrt{30} m_{1}-60 \sqrt{30} m_{2}+\tilde{s} (-127 \lambda_{2}+173 \lambda_{3}+200 \lambda_{4}+140 \lambda_{5})\right)$ \\
    $(\mathbf{144},-3)$ &
     $\mathbb{C}$&
    $\tfrac{1}{96} \tilde{s} \left(12 \sqrt{30} (m_{1}+m_{2})-\tilde{s} (52 \lambda_{2}-32 \lambda_{3}+76 \lambda_{4}+\lambda_{5})\right)$ \\
    $(\mathbf{210},+0)$ &
     $\mathbb{R}$&
    $\tfrac{1}{240} \tilde{s} \left(3 \sqrt{30} (13 m_{1}+4 m_{2})-4 \tilde{s} (31 \lambda_{2}+\lambda_{3}+7 (4 \lambda_{4}+\lambda_{5}))\right)$ \\
    \bottomrule
\end{tabular}
\end{center}
\end{table}

Considering the labeling in the tables: when there is more than one copy of the representation $R$ of subgroup $G$ in the $\mathbf{650}$, we distinguish the eigenstates by labeling them with a numeric index. This happens only for group $G_{333}$ in Table~\ref{tab:masses-vacuum-su3su3su3}.   

For a given parameter point $(M^{2},m_{k},\lambda_{l})$, the VEV solution associated to the symmetry $G$ is a local minimum of the potential $V_{650}$ if all the masses are non-tachyonic for that case.\footnote{Note that in a realistic model, other scalar representations would be needed beside the $\mathbf{650}$. The newly introduced  states would also need to be non-tachyonic, but since their mass expressions would necessarily involve only new parameters, that analysis is of no relevance to the parameters in $V_{650}$ considered here regardless of the scalar sector extension.}
We analyze the non-tachyonic regions of parameter space numerically in Section~\ref{sec:global-minima}, but conclude the discussion here with some observations on the analytic form of mass expressions:
\begin{itemize}
	\item 
	A vanishing mass expression in Tables~\ref{tab:masses-vacuum-su3su3su3}--\ref{tab:masses-vacuum-so10u1} corresponds to the would-be Goldstone states, as confirmed by the lists of broken generators denoted by non-underlined terms in the decomposition of $\mathbf{78}$ found in Table~\ref{table:E6-maximal-subgroups}. Note that this adjoint $\mathbf{78}$ is taken as a real representation, so the complex representations in the decomposition have their degrees of freedom listed twice --- both in the original form and as a complex conjugate. In particular, this remark applies for the $(\mathbf{3},\mathbf{\bar{3}},\mathbf{\bar{3}})$ of $G_{333}$ and the $(\mathbf{16},-3)$ of $G_{10,1}$.\par
	The correct would-be Goldstone structure is a non-trivial consistency check of our vacuum and mass computation. 
	\item The masses-square have a structure consistent with power-counting of mass dimensions. The only dimensionful quantities in the theory are $m_{1}$, $m_{2}$ and $M^{2}$, with the the latter being absorbed into the VEV $s$ or $\tilde{s}$ in the mass expressions. The mass-square expressions can thus be formed only from $7$ linearly independent terms: $m_{k}\tilde{s}$ ($k=1,2$) arising from cubic terms in the scalar potential, and $\lambda_{l}\tilde{s}^{2}$ (for $l=1,\ldots,5$) arising from quartic terms. For the $G_{333}$ case, the same applies if we replace $\tilde{s}$ with $s$ in the argumentation. \par
	The explicit forms in the tables confirm this analytic structure. Furthermore, the masses within each table are linearly independent; the exception is the case $G_{333}$, where $9$ different mass-square expressions span a $7$-dimensional linear space, implying $2$ independent mass relations. Specifically, the $2$ mass sum rules in the trinification case can be written as
	\begin{align}
	2 \,m^{2}_{(\mathbf{8},\mathbf{1},\mathbf{1})} +
	8 \,m^{2}_{(\mathbf{8},\mathbf{8},\mathbf{1})} &=
	m^{2}_{(\mathbf{1},\mathbf{1},\mathbf{1})_{2}} +
	9\,m^{2}_{(\mathbf{6},\mathbf{3},\mathbf{3})},\\
	m^{2}_{(\mathbf{1},\mathbf{8},\mathbf{8})} +
	9\, m^{2}_{(\mathbf{3},\mathbf{\bar{3}},\mathbf{\bar{3}})_{1}} +
	3\, m^{2}_{(\mathbf{6},\mathbf{3},\mathbf{3})} &=
	2\, m^{2}_{(\mathbf{1},\mathbf{1},\mathbf{1})_{2}} +
	2\, m^{2}_{(\mathbf{1},\mathbf{8},\mathbf{1})} +
	3\, m^{2}_{(\mathbf{3},\mathbf{6},\mathbf{\bar{3}})} +
	6\, m^{2}_{(\mathbf{8},\mathbf{8},\mathbf{1})}.	
	\end{align}
	\item All $3$ vacua in the trinification case $G_{333}$ are equally suitable, i.e.~have the same potential value. We chose the LR-symmetric case with $a=0$, which then manifests in a LR-symmetric scalar spectrum, as seen from mass degeneracies in Table~\ref{tab:masses-vacuum-su3su3su3}. The spectra of the other two vacua consist of the same masses reshuffled among the representations, in accordance with the action of the parity that transforms the LR-symmetric vacuum to the desired one. \par
    The set of $3$ solutions in the trinification case also indicates a possible formation of domain walls during the first stage of symmetry breaking. Also, $\mathbb{Z}_{2}$-strings are the other type of possible topological defect arising due to preserved parity, as has been pointed out for the $D$-parity trinification case in Ref.~\cite{Chakrabortty:2019fov}. We don't purse here the question of topological defect further; they are not problematic for cosmology if the scale of inflation is lower than the breaking scale. 
	\item Curiously, the mass of the second trinification singlet $(\mathbf{1},\mathbf{1},\mathbf{1})_{2}$ in Table~\ref{tab:masses-vacuum-su3su3su3} is a linear combination of the massive parameters $m_{k}$, and hence vanishes if one takes the limit $m_{k}\to 0$.\par
	This can be understood directly from the restricted potential $V(s,a)$. Taking $m_{k}=0$ removes the cubic terms, and the remaining quadratic and quartic terms are functions only of the combination $s^{2}+a^{2}$. In this limit the restricted potential $V$ thus becomes rotationally symmetric in the $(s,a)$ plane, and the vacuum manifold of $V$ becomes a circle, i.e.~$1$-dimensional, on which all $P_{i}$ for $i$ from $1$ to $6$ lie. 
	The second $G_{333}$-singlet mass eigenstate can thus be interpreted as a pseudo-Goldstone boson associated to the explicit breaking of the $\SO(2)$ rotational symmetry in the $(s,a)$-plane by cubic terms with parameters $m_{k}$. Note, however, that in the $m_{k}\to 0$ limit only the restricted potential $V$ in the $G_{333}$-singlet plane becomes rotationally symmetric, and not the full potential $V_{650}$. This is not surprising, since the aforementioned $\SO(2)$ symmetry belongs to a $\SO(650)$ group of transformations of the real representation $\mathbf{650}$, but not to the smaller $\mathrm{E}_{6}$. In particular, the $M^{2}$ and $\lambda_{1}$ terms are symmetric under this $\SO(2)$, while $\lambda_{l}$ for $l=2,3,4,5$ explicitly break it in the full space and preserve it only in the $(s,a)$-plane. 
\end{itemize}

\subsection{Candidates for global minima \label{sec:global-minima}}

As a final step in our vacuum analysis, we (partially) address the question of the global minimum of the potential $V_{650}$ in Eq.~\eqref{eq:explicit-potential}. 
This is achieved by comparing the value of the potential $V_{650}$ of the solutions to vacua corresponding to the five ``relevant'' cases of Table~\ref{table:E6-maximal-subgroups} that we found in Section~\ref{sec:analysis-of-vacua}, and we refer to the deepest among them as the ``global'' minimum.

We now proceed as follows: the procedure for determining the ``global'' minimum for a given parameter point and the associated caveats are discussed in
Section~\ref{sec:global-procedure}, while concrete numerical results are then given in Section~\ref{sec:global-results}.

\subsubsection{Procedure and limitations \label{sec:global-procedure}}

First, let us specify the procedure by which a parameter point $(M^{2},m^{2}_{k},\lambda_{l})$ is assessed, in part summarizing the results obtained from local analyses. We perform the following computational steps:
\begin{enumerate}
	\item The potential $V_{650}$ must be bounded from below. We test this only in the singlet directions; in particular, we confirm that $\sum_{i}\beta_{i}\lambda_{i}>0$ for all cases of $\vec{\beta}$ in Table~\ref{table:vacua-alfa-beta}. If this criterion is not passed, we do not consider the parameter point as viable.
	\item All VEV solutions for the five ``relevant'' breaking possibilities in Table~\ref{table:E6-maximal-subgroups} are determined from Eq.~\eqref{eq:sa-solution-1}, where $m$ and $\lambda$ are determined from Eq.~\eqref{eq:parameters} and Table~\ref{table:vacua-alfa-beta}.
	\item Each VEV solution, together with its associated values of $m$ and $\lambda$, is inserted back into $V$ of Eq.~\eqref{eq:singlet-potential-1} to determine the corresponding values of the potential in the candidate points. Additionally, the potential value for the stationary point with a vanishing VEV $\langle \mathbf{650}\rangle=0$, corresponding to the unbroken $\mathrm{E}_{6}$ phase, is considered.
	\item For the regular cases of Table~\ref{table:E6-maximal-subgroups}, which contain the SM group and are thus the physically relevant ones, we compute the masses for each of their VEV solutions using Tables~\ref{tab:masses-vacuum-su3su3su3}--\ref{tab:masses-vacuum-so10u1} and check for possible tachyonicity. We thus unambiguously determine which cases correspond to local minima.
	\item The case with the lowest value of the potential that is also a determined to be a local minimum is then considered the sole candidate for the ``global'' minimum. It is considered a viable ``global'' minimum if it also passes the following two physically motivated constraints: 
        \begin{itemize}
            \item[(i)]
            The cubic couplings $m_{1}$ and $m_{2}$ should not be too large compared to the heaviest scalar mass (computed in the ``global'' minimum candidate only), otherwise the scattering amplitudes become non-unitary at low energies, see e.g.~\cite{Staub:2018vux}.
            \item[(ii)]
            A similar limitation applies to the gauge boson masses in the ``global'' minimum candidate: they should not be much larger than the heaviest scalar state lest the tree-level analysis becomes threatened by large one-loop corrections to the potential~\cite{Coleman:1973jx}. 
        \end{itemize}
    For both criteria, the ``much heavier'' threshold is taken to be a factor $10$ at the level of masses-square, and is explicitly implemented as the following simple criteria:\footnote{In practice, the criteria of \eqref{eq:physical-criteria} are unlikely to be violated for a random parameter point; their exact 
    formulation is thus not crucial, and we essentially use them as a very simple way of confirming that the breaking phenomenon occurs at roughly a single energy scale, which is not destabilized by any of the massive parameters or VEVs at the quantum level.}
     \begin{align}
        \max\{m_{1}^{2},m_{2}^{2}\}&\leq 10\,\max_{R}\, m^{2}_{R},&
        m^{2}_{\text{gauge}}&\leq 10\,\max_{R}\, m^{2}_{R}, \label{eq:physical-criteria}
    \end{align}
    where $R$ goes over the irreducible scalar representations of $G$ (the residual symmetry in the ``global'' minimum candidate) present in the $\mathbf{650}$, and $m^{2}_{\text{gauge}}$ labels the gauge boson mass from Table~\ref{table:vacua-alfa-beta} when breaking into $G$ with the unified gauge coupling value taken as $g\simeq 0.5$.
\end{enumerate}
The above procedure allows to asses any point in parameter space and determine its ``global'' minimum, which in turn serves as our best estimate for the actual global minimum of the scalar potential. The procedure, however, does have some caveats:
\begin{itemize}
    \item Local minima that are lower than the ``global'' one may in principle exist, since we did not fully classify all VEV solutions in the $650$-dimensional space of scalars.
    As discussed in Section~\ref{sec:maximal-subgroups}, however, the only other plausible minima consistent with Michel's conjecture (assuming discrete remnants of $\mathrm{E}_{6}$ are not present) lead to subgroups of the last three cases of Table~\ref{table:E6-maximal-subgroups}. Indeed, the representation $\mathbf{650}$ contains singlets under all maximal subgroups of the three cases; we list all these candidates in Table~\ref{tab:subgroups-level2}. Among them, the candidates consistent with Michel are those who are not simultaneously subgroups of the ``relevant'' cases (up to conjugation). \par
    We do not proceeded further with this analysis, but one can argue that these small groups are not favored to produce a local minimum in most of the parameter space:
    under small subgroups of $\mathrm{E}_{6}$, the $\mathbf{650}$ decomposes into a large number of irreducible representations, leading to a large set of non-tachyonicity conditions that would all need to simultaneously hold.
    \item The ``global'' analysis searches for the global minimum of the potential among the local minima. Consider though the well-known counterexample of a ``runaway'' potential in two dimensions:  
	\begin{align}
		V(x,y)&=(M^2-xy)^2+m^2\,x^2,
	\end{align}	
where $m,M>0$. Clearly $V\geq 0$. The only stationary point is $(0,0)$ and it is a saddle point, the value $V=0$ cannot be reached by any $(x,y)$, while for any $\epsilon>0$ the point $(x_{\epsilon},y_{\epsilon})=(m\sqrt{\epsilon},\frac{M^2}{m\sqrt{\epsilon}})$ gives the potential value $V(x_\epsilon,y_\epsilon)=\epsilon\,m^{4}$, which is arbitrarily close to zero for a small enough $\epsilon$. This potential is bounded from below, but does not have an absolute minimum at any point. \par
    The existence of the global minimum for a continuous potential is guaranteed  only in a compact domain, and the key feature of the counterexample is that the potential does not diverge at infinity in all directions. Such a pathology though is unlikely to arise in our $V_{650}$ case for the generic parameter point, since already the quartic invariant associated to the parameter $\lambda_{1}$ in Eq.~\eqref{eq:explicit-potential} diverges at infinity for all directions of $\mathbf{X}$.
    \begin{table}[htb]
	\centering
	\caption{The maximal subgroups of groups $G$, which are in turn maximal subgroups of $\mathrm{E}_{6}$. We listed only those $G$ which don't already have a $G$-invariant state in the $\mathbf{650}$. \label{tab:subgroups-level2}}
	\begin{tabular}{ll}
		\toprule	
		$G$&maximal subgroups of $G$\\
		\midrule
		$G_{2}$& $\SU(3)$, $\SU(2)\times\SU(2)$, $\SU(2)$\\
		$\SU(3)$& $\SU(2)\times\UU$, $\SU(2)$\\
		$\mathrm{Sp}(8)$& $\SU(4)\times\UU$, $\SU(2)\times\mathrm{Sp}(6)$,
			$\mathrm{Sp}(4)\times\mathrm{Sp}(4)$, $\SU(2)$, $\SU(2)^3$\\
		\bottomrule
	\end{tabular}
	\end{table}
    \item Boundedness from below was checked only in the $5$ main VEV directions, i.e.~$\lambda >0$ for all cases in Table~\ref{table:vacua-alfa-beta}. This is only a necessary condition that the full potential $V_{650}$ is bounded from below.
	\par 
    Formulating the most general sufficient conditions for boundedness from below in the full $650$-dimensional space is a difficult task. Although boundedness can be checked in some special cases, e.g.~in the case of an adjoint of $\SU(5)$~\cite{MacKenzie:1982fi}, we are not aware of a general procedure. 
    We do make the observation, however, that the invariants with coefficients $\lambda_{1}$ and $\lambda_{2}$ in Eq.~\eqref{eq:explicit-potential} give only positive directions for quartic terms, since $\mathbf{X}$ is a Hermitian matrix. Therefore, qualitatively, one sufficient condition for boundedness from below is to have $\lambda_{1},\lambda_{2}>0$ dominating over the other quartic terms, i.e.~$\lambda_{1,2}\gg |\lambda_{3,4,5}|$. 
	\par 
	Quantitatively, the argument can be pushed a bit further by noting that the other quartic invariants involve the invariant tensors $d$ and $D$, 
	which consist only of numeric values $\pm 1$ or $0$, see Appendix~\ref{appendix:E6}. It thus seems likely that a condition in the form
	\begin{align}
		\lambda_{1}+\lambda_{2}	> C\,\left(|\lambda_{3}|+|\lambda_{4}|+|\lambda_{5}|\right)
	\end{align}
	would be sufficient for boundedness from below for some unknown positive constant $C\sim \mathcal{O}(1)$.
 \item The entire analysis is performed only at tree-level. Quantum corrections to the effective scalar potential may be large, especially considering the large number of fields that can run in the loops. This consideration may considerably shrink the range of couplings in which the computation is amenable to perturbative methods, see e.g.~\cite{Jarkovska:2021jvw,Jarkovska:2023zwv} for a recent example from $\SO(10)$ GUT.
	\par
	We do not pursue here this issue further, and simply limit the region for dimensionless parameters to $|\lambda_{l}|\leq 1$.
\end{itemize}

Keeping in mind the limitations discussed above, the analysis of the ``relevant'' cases is nevertheless expected to yield the best candidate for the global minimum.

\subsubsection{Results for ``global'' minima \label{sec:global-results}}
 
We now investigate the parameter space $(M^{2},m_{k},\lambda_{l})$ for ``global'' minima with a symmetry among the ``relevant'' cases in accordance with the considerations of Section~\ref{sec:global-procedure}. The parameters are those present in $V_{650}$ of Eq.~\eqref{eq:explicit-potential}.
 
\def\Yes{\checkmark}
\def\No{-}
\begin{table}[htb]
	\centering
	\caption{A list of benchmark points. We provide for each a label, the parameter values $(M^{2},m_{k},\lambda_{l})$, the existence of local minima for regular cases ($\Yes$ for yes and $\No$ for no), and the symmetry in the ``global'' minimum. \label{tab:benchmark-points}}
	\begin{tabular}{c@{\quad}rrr@{\ \;}rrrr@{\ \;}r@{\quad}c@{\,}c@{\,}c@{\quad}l}
		\toprule
		point & 
		$M^{2}$ & $m_{1}$& $m_{2}$ & 
		$\lambda_{1}\phantom{0}$ & $\lambda_{2}$ & 
		$\lambda_{3}$ & $\lambda_{4}$ & $\lambda_{5}\phantom{0}$ 
		& $G_{10,1}$ & $G_{62}$ & $G_{333}$ & ``global'' minimum \\
		\midrule
		A & $-1$ & $1.0$ & $0.0$ & 
			$0.2\phantom{0}$ & $0.0$ & $0.2$ & $0.0$ & $0.0\phantom{0}$ & 
			$\Yes$ & $\No$ & $\No$ & $\SO(10)\times\UU$\\
		B & $-1$ & $0.0$ & $1.0$ & 
			$0.2\phantom{0}$ & $0.0$ & $0.0$ & $0.0$ & $0.0\phantom{0}$ & 
			$\No$ & $\Yes$ & $\No$ & $G_{62}$\\
		C & $-1$ & $0.5$ & $1.0$ & 
			$0.15$ & $0.2$ & $0.2$ & $0.2$ & $0.15$ & 
			$\No$ & $\No$ & $\Yes$ & $G_{333}$\\
		D & $-1$ & $1.0$ & $1.0$ & 
			$0.1\phantom{0}$ & $0.0$ & $0.0$ & $0.0$ & $0.0\phantom{0}$ & 
			$\No$ & $\Yes$ & $\No$ & $\mathrm{F}_{4}$\\
		E & $-1$ & $1.0$ & $1.0$ & 
			$0.5\phantom{0}$ & $0.6$ & $0.0$ & $0.2$ & $0.0\phantom{0}$ & 
			$\Yes$ & $\Yes$ & $\No$ & $\SU(3)\times\mathrm{G}_{2}$\\
		F & $-1$ & $0.6$ & $0.7$ & 
			$0.3\phantom{0}$ & $0.0$ & $0.3$ & $0.2$ & $0.4\phantom{0}$ & 
			$\Yes$ & $\Yes$ & $\Yes$ & $\SO(10)\times\UU$\\[5pt]
		\bottomrule	
	\end{tabular}
\end{table}

First, we provide a list of benchmark points in Table~\ref{tab:benchmark-points} that serves as an existence proof for various scenarios. The points were carefully selected so that the following remarks can be made: 
\begin{itemize}
	\item Notice that points A--E cover all the symmetry cases for a global minimum we considered from Table~\ref{table:E6-maximal-subgroups}. 
	\item Points A--C correspond to situations, where the global minimum breaks to one of the regular subgroups. For each of these there are no local minima of the potential corresponding to the other two regular cases, since they are destabilized by tachyonicity and the associated stationary points 
    are saddle points of the full potential $V_{650}$. 
	\item Case F is an example where a local minimum exists for all $3$ types of regular cases. The $\SO(10)\times\UU$ minimum is the deepest among them.
	\item Points D and E break into one of the special cases, where only the depth of the minimum was checked and not the masses. Although local minima to regular cases exist for these points, they are metastable.
\end{itemize}

Second, we perform a search for points where the ``global'' minimum is in one of the regular scenarios of Table~\ref{table:E6-maximal-subgroups}, i.e.~one of the phenomenologically interesting cases. 
Although a local minimum with a long enough vacuum lifetime would already be phenomenologically viable, 
the demand for a ``global'' minimum ensures exclusivity of each parameter point and thus yields a phase diagram in the $8$-dimensional space $(M^{2},m_{k},\lambda_{l})$. 

We limit our search to the parameter region 
\begin{align}
    |M^{2}|&\leq 1,& 
    |m_{k}|&\leq 1,& 
    |\lambda_l|&\leq 1,
    & m_{1}&\geq 0. \label{eq:input-parameter-ranges}
\end{align}
The limitation on the dimensionless couplings $\lambda_l$ is essentially a rough perturbativity bound, cf.~Section~\ref{sec:global-procedure}, while $m_1$ can be made non-negative 
by redefining the sign of $\mathbf{X}$ in $V_{650}$ if necessary.

Note that the ``global'' analysis does not change qualitatively under mass rescaling,
i.e.~under \hbox{$m_{k}\mapsto \eta\, m_{k}$} and \hbox{$M^{2}\mapsto \eta^2\, M^{2}$}, so there is a one parameter redundancy in the space $(M^{2},m_{k},\lambda_{l})$. When $M^{2}\neq 0$, one could in principle rescale the dimensionful parameters so that $M^{2}=\pm 1$ and eliminate the redundancy, but a separate analysis is then required for each discrete case. Since this obscures the smooth transition in quantitative results when $M^{2}$ changes sign, we rather perform the analysis in the full parameter space. In this context, the limits on $M^{2}$ and $m_{k}$ in Eq.~\eqref{eq:input-parameter-ranges} are not important, provided they define a region with equal opportunity for either $M^{2}$ or $m_{k}$ to dominate.

For each regular case, we scan for a dataset of $10^4$ viable parameter points. The search was performed by first finding a starting generation of points in each case, and then obtaining new ones by use of a stochastic variant of the differential evolution algorithm~\cite{diffevolalg} (version ``DE/rand/1''~with a random choice of $F \in (0.5,2)$ for each point).

To present the results visually, the points must be projected from the 8D space $(M^{2},m_{k},\lambda_{l})$ down to 1D or 2D for plotting.  Colors encode the different datasets: 
\textbf{\textcolor{myRed}{red}} for trinification $\SU(3)^3$ (left bar),
\textbf{\textcolor{myBlue}{blue}} for $\SU(6)\times\SU(2)$ (middle bar),
and \textbf{\textcolor{myGreen}{green}} for $\SO(10)\times\UU$ (right bar).

\begin{figure}[htb]
    \centering
    \includegraphics[height=6.3cm]{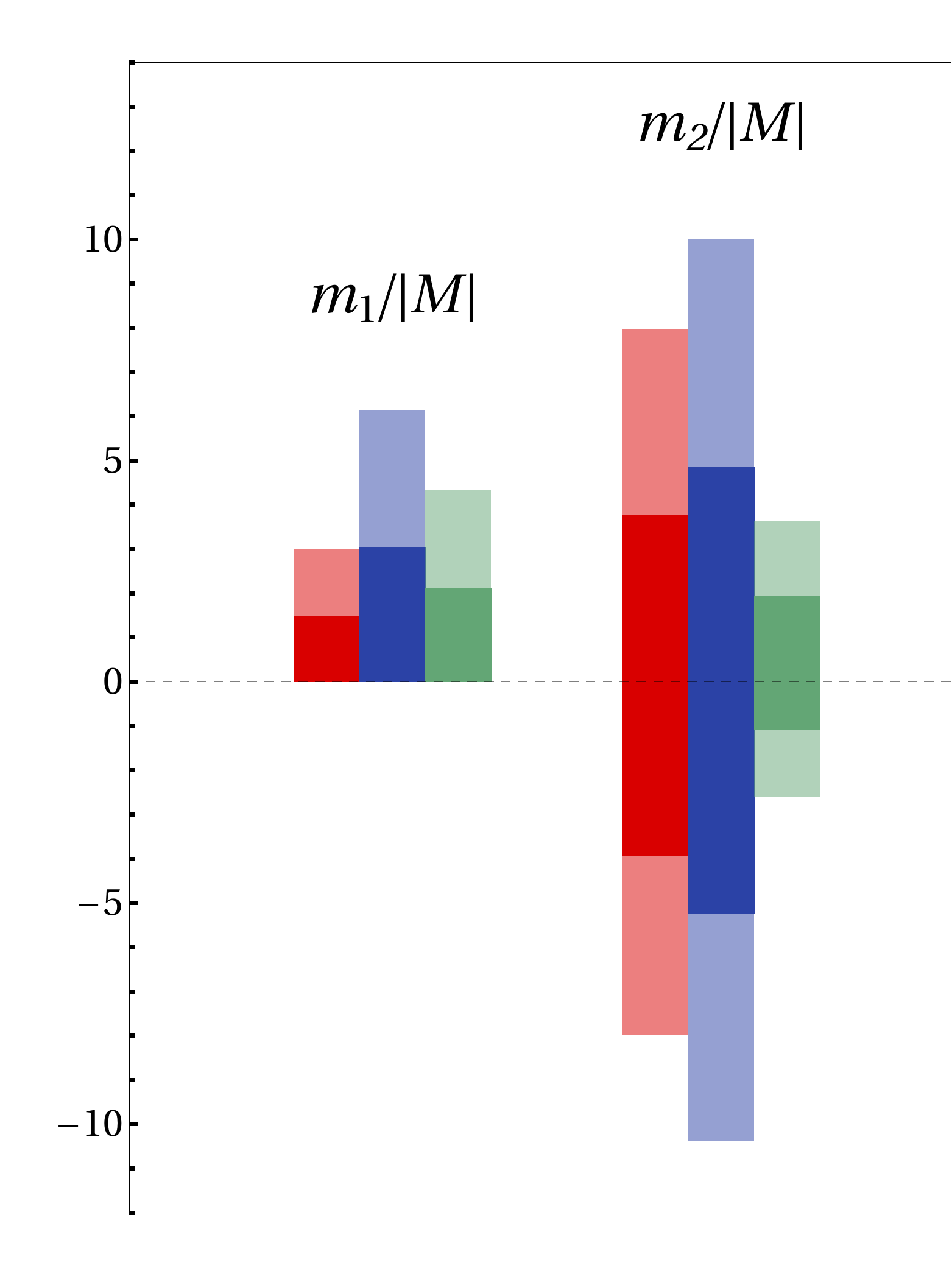}
    \includegraphics[height=6.3cm]{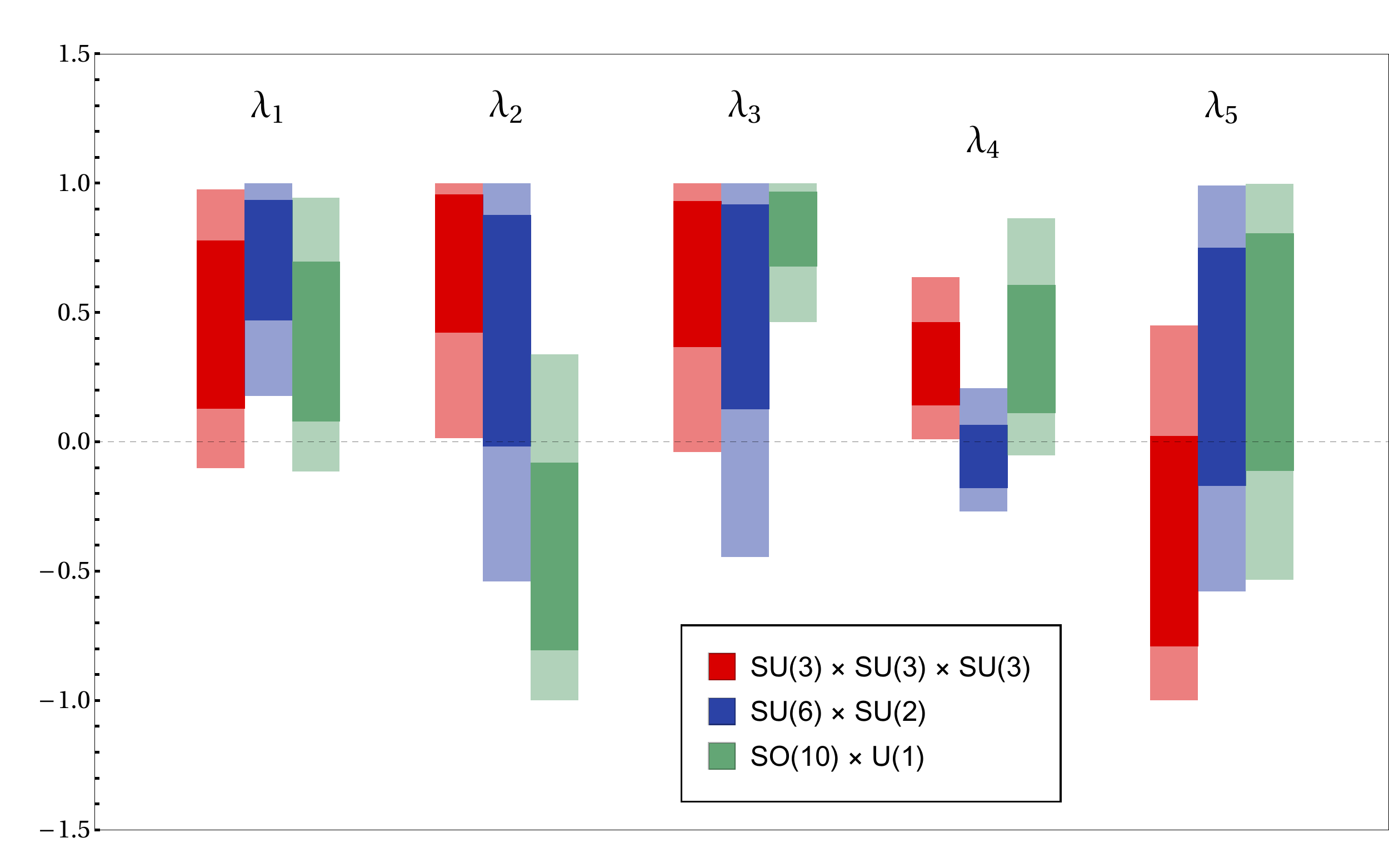}
    \caption{The $1$- and $2$-$\sigma$ HDI ranges (with decreasing opacity) for the massive parameters $m_{k}$ and the dimensionless couplings $\lambda_k$. The colors denote the different cases of the resulting global minimum, as shown in the legend. \label{fig:HDI-ranges}}
\end{figure}

We show the 1D \textit{highest density intervals} (HDIs) for separate input parameters in Figure~\ref{fig:HDI-ranges}. Due to mass-scale redundancy, we show only dimensionless ratios of mass parameters. Also, $m_{1}/|M|>0$ in accordance with Eq.~\eqref{eq:input-parameter-ranges}. 
The vertical bars indicate the $1$- and $2$-$\sigma$ HDIs in decreasing opacity.

\begin{figure}[htb]
    \centering
    \includegraphics[width=7.5cm]{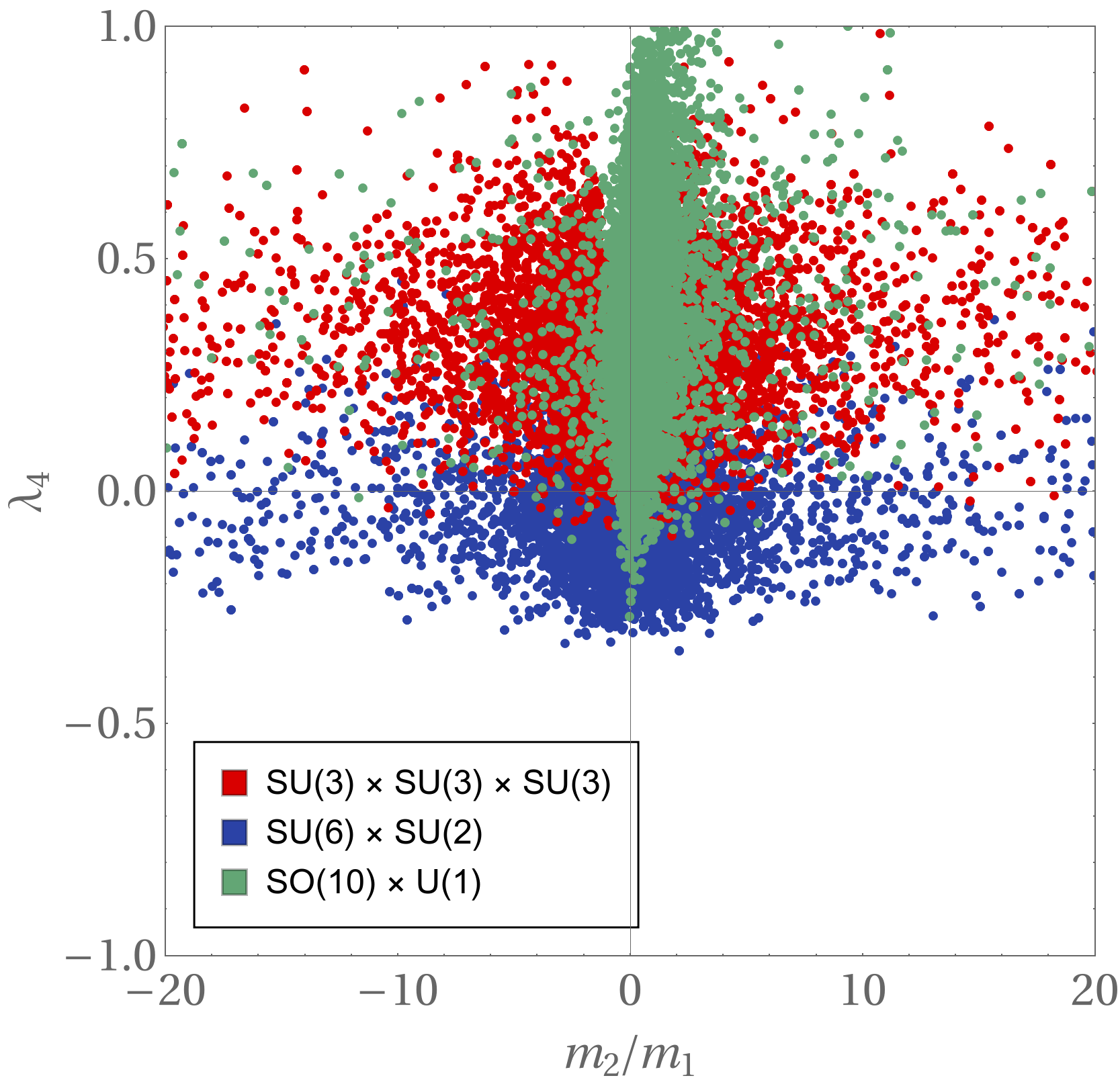}
    \includegraphics[width=7.5cm]{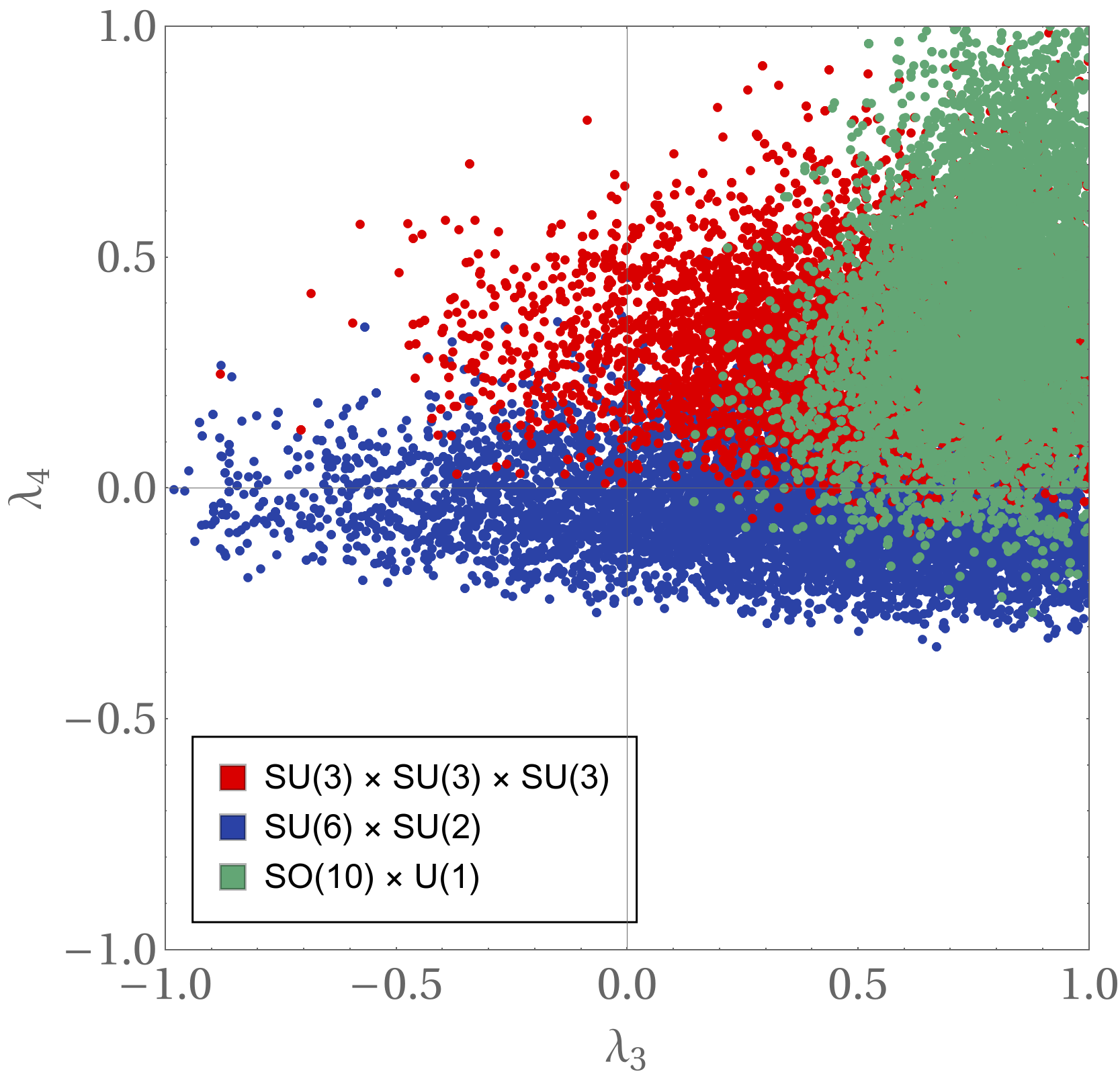}
    \caption{Correlation scatter plots in the planes $(m_{2}/m_{1})$-$\lambda_{4}$ (left panel) and $\lambda_{3}$-$\lambda_{4}$ (right panel). The colors denote different cases, see the legends. \label{fig:scatter-plots}}
\end{figure}

Additionally, it is also instructive to show 2D correlation plots for specific pairs of inputs. Figure~\ref{fig:scatter-plots} shows scatter plots in the $(m_{2}/m_{1})$-$\lambda_{4}$ and $\lambda_{3}$-$\lambda_{4}$ planes.
These choices of pairs best discriminate between the various regular cases.

The results from Figures~\ref{fig:HDI-ranges} and~\ref{fig:scatter-plots} show the (projected) regions, where parameter values most often lead to a given type of ``global'' minimum. One can make the following observations:
\begin{itemize}
	\item ``Global'' minima with $G_{333}$ or $G_{62}$ symmetry prefer large ratios $|m_{2}/m_{1}|$, while $G_{10,1}$ prefers small such ratios, large positive values of $\lambda_{3}$ and negative values of $\lambda_{2}$.
	\item The parameter $\lambda_{4}$ is best for discriminating ``globally'' between the $G_{333}$ and $G_{62}$ cases.  
	\item There is a high preference for $\lambda_{1}>0$ in all cases, and $\lambda_{3}$ and $\lambda_{4}$ also prefer positive ranges. This is likely due to the conditions for boundedness from below of the full potential $V_{650}$, cf.~Section~\ref{sec:global-procedure}.
\end{itemize}

In summary, we see from Table~\ref{tab:benchmark-points} that ``global'' minima for all ``relevant'' cases  exist somewhere in parameter space, while Figures~\ref{fig:HDI-ranges} and \ref{fig:scatter-plots} reveal for each regular case the preferred (projected) regions of parameters.

\section{Conclusions \label{sec:conclusions}}

Unlike the smaller GUT symmetries $\SU(5)$ or $\SO(10)$, the group $\mathrm{E}_{6}$ offers some unique opportunities in GUT model building, namely a breaking chain to the SM group through $\SU(3)^3$ (trinification) or $\SU(6)\times\SU(2)$. The real representation $\mathbf{650}$ of $\mathrm{E}_{6}$ is the minimal and only realistic way to implement these breaking chains, since the next-largest representation of $\mathrm{E}_{6}$ with $G_{333}$- or $G_{62}$-singlets is of dimension $2430$. A Yang-Mill theory with only a real scalar $\mathbf{650}$ is still asymptotically free, while the presence of $\mathbf{2430}$ leads inextricably to a Landau pole within an order of magnitude.

We investigated in this paper the most general renormalizable scalar potential with a single copy of the representation $\mathbf{650}$, which consists of one quadratic, two cubic and five quartic linearly independent invariants. We obtained explicit VEV solutions that break $\mathrm{E}_{6}$ to the following cases:
\begin{itemize}
	\item[(i)] the regular maximal subgroups $\SU(3)^3$, $\SU(6)\times\SU(2)$ and $\SO(10)\times\UU$,
	\item[(ii)] the special maximal subgroups $\mathrm{F}_{4}$ and $\SU(3)\times\mathrm{G}_{2}$.
\end{itemize}
Although this may not be a complete classification of solutions, we argued that the above list consists of the most important cases, at least in most of parameter space. 

Phenomenologically, the important cases are those with a regular maximal subgroup, since these still retain the SM group as a subgroup. For these vacuum solutions we computed the scalar spectrum, thus studying when they correspond to a local minimum of the full scalar potential. In a (partial) global analysis, we then scanned the parameter space for points which support each of these regular cases as the lowest minimum (among those analyzed); indeed, non-empty regions 
were identified for all of them, effectively producing a phase diagram withing the $8$-dimensional parameter space.

An aspect of the trinification solution worth emphasizing is that it also preserves a discrete part of $\mathrm{E}_{6}$, analogous to $D$-parity in $\SO(10)$: either LR, CR or CL parity remain unbroken, where intuitively these parities exchange the factors of $\SU(3)_C\times\SU(3)_L\times\SU(3)_{R}$. The imprint of the remaining parity is manifest in the spectrum, and it may lead to topological defects, in particular domain walls and strings. 

Finally, the representation $\mathbf{650}$ has a an intriguing outlook in $\mathrm{E}_{6}$ model building, especially in the non-supersymmetric case where the breaking is expected to occur in multiple stages. This paper addressed the possibilities for the first stage in $\mathrm{E}_{6}$ breaking without further regard to additional phenomenological considerations or a concrete model setup; it serves as a piece in the wider effort of $\mathrm{E}_{6}$ GUT model building, including our future work.

\section*{Acknowledgments}

The work of KSB is supported in part by the U.S.~Department of Energy under grant number
DE-SC0016013. BB acknowledges the financial support from the Slovenian Research Agency (research core funding No. P1-0035 and in part research grant J1-4389). VS acknowledges financial support from the Grant Agency of the Czech Republic (GA\v{C}R) via Contract No.~20-17490S, as well as from the Charles University Research Center Grant No.~UNCE/SCI/013. We thank Raymond Volkas for 
discussion and pointing out Ref.~\cite{Lonsdale}.

\appendix

\section{The special embeddings of $\mathrm{F}_{4}$ and $\SU(3)\times\mathrm{G}_{2}$ into $\mathrm{E}_{6}$\label{appendix:special-embeddings}}

We provide in this Appendix some technical details on dealing with the Lie algebra of $\mathrm{E}_{6}$, as well as the special embeddings of $\mathrm{F}_{4}$ and $\SU(3)\times\mathrm{G}_{2}$.

Note that a special embedding $H\subset G$ breaks the rank of the larger group $G$.
Suppose we have a basis of generators in $G$, such that every generator has well-defined quantum numbers of $G$ (this is the basis of lowering and raising operators). Suppose we have an analogous basis of generators in $H$, such that each of those have well-defined quantum numbers in $H$. Due to the rank-breaking property of the maximal embedding, the basis of $H$ cannot be simply a subset of the generators of $G$. Instead, some generators of $H$ will necessarily have to be linear combinations of those in $G$ --- a feature we shall indeed see explicitly in the embeddings of $\mathrm{F}_{4}$ and $\SU(3)\times\mathrm{G}_{2}$. This makes the analysis of special embeddings more complicated, hence its relegation to the appendix.

The most convenient treatment of the exceptional groups $\mathrm{E}_{6}$, $\mathrm{F}_{4}$ and $\mathrm{G}_{2}$ is perhaps by making use of their maximal regular subgroups $\SU(3)^n$, see Table~\ref{table:exceptional-groups}. We describe each of these cases in turn, and specify explicit embeddings in the following subsections.

\begin{table}[htb]
\centering
\caption{A comparison of exceptional groups $\mathrm{E}_{6}$, $\mathrm{F}_{4}$ and $\mathrm{G_{2}}$, and the branching rules for their adjoints to a maximal $\SU(3)^n$ in each case, which facilitates a convenient language for their description. \label{table:exceptional-groups}}
\vspace{-0.5cm}
\begin{tabular}{rr@{\qquad}ll}
	\toprule
	$G$&adjoint&decomposition&maximal $\SU(3)^n\subset G$\\
	\midrule
	$\mathrm{E}_{6}$&$\mathbf{78}$&
	$\mathbf{8}_C+\mathbf{8}_L+\mathbf{8}_R
	+(\mathbf{3},\mathbf{\bar{3}},\mathbf{\bar{3}})
	+(\mathbf{\bar{3}},\mathbf{3},\mathbf{3})$	
	&$\SU(3)^{3}\quad\equiv\quad \SU(3)_C\times\SU(3)_L\times\SU(3)_{R}$\\[6pt]
	$\mathrm{F}_{4}$&$\mathbf{52}$&
	$\mathbf{8}_C+\mathbf{8}_{LR}
	+(\mathbf{3},\mathbf{\bar{6}})
	+(\mathbf{\bar{3}},\mathbf{6})$	
	&$\SU(3)^{2}\quad\equiv\quad \SU(3)_C\times\SU(3)_{LR}$\\[6pt]
	$\mathrm{G}_{2}$&$\mathbf{14}$&
	$\mathbf{8}+\mathbf{3}+\mathbf{\bar{3}}$	
	&$\SU(3)^{1}$\\
\bottomrule
\end{tabular}
\end{table}

\subsection{The algebra of $\mathrm{E}_{6}$ \label{appendix:E6}}

As already stated in Section~\ref{sec:e6-generators}, the $78$ generators of $\mathrm{E}_{6}$ can be labeled by
\begin{align}
	(T_{C})^{A},\quad
	(T_{L})^{A},\quad
	(T_{R})^{A},\quad
	t^{\alpha}{}_{aa'},\quad
	\bar{t}_{\alpha}{}^{aa'}, 
\end{align}
where $A$ is the adjoint $\SU(3)$ index (for the appropriate factor), while $\{\alpha,a,a'\}$ are fundamental for $\SU(3)_{\{C,L,R\}}$. The most general element in the Lie algebra $\mathfrak{e}_{6}$ is thus written as
\begin{align}
    \alpha_C{}^{A}\,T_{C}^{A}+
    \alpha_L{}^{A}\,T_{L}^{A}+
    \alpha_R{}^{A}\,T_{R}^{A}+
    \beta_{\alpha}{}^{aa'}\,t^{\alpha}{}_{aa'}
    +\beta^{\ast\,\alpha}{}_{aa'}\,\bar{t}_{\alpha}{}^{aa'},
\end{align}
where the $24$ coefficients $\alpha_C{}^{A}$, $\alpha_L{}^{A}$, $\alpha_R{}^{A}$ are real, while the $27$ coefficients $\beta_{\alpha}{}^{aa'}$ are complex.
For quick reference by ``sector'', we shall sometimes suppress the indices and simply refer to the generators as $\{T_{C},T_{L},T_{R},t,\bar{t}\}$. 

The explicit $\mathrm{E}_{6}$ commutation relations are then written as

\def\TC{T_C}
\def\TR{T_R}
\def\TL{T_L}
\def\T{{t}}
\def\TB{\bar{t}}

\begin{align}
\big[ \TC^A,\TR^B \big]= \big[\TR^A,\TL^B \big]= \big[\TL^A,\TC^B \big]&=0, \label{eq:commutation-e6-begin}
\end{align}
\begin{align}
\big[ \TC^A,\TC^B \big]&=i f^{ABD} \;\TC^D,\\
\big[ \TL^A,\TL^B \big]&=i f^{ABD} \;\TL^D,\\
\big[ \TR^A,\TR^B \big]&=i f^{ABD} \;\TR^D,
\end{align}
\begin{align}
\big[ \TC^A,\T^{\alpha}{}_{aa'} \big]&= -\tfrac{1}{2}(\lambda^A)^\alpha{}_\beta \;\T^{\beta}{}_{aa'},\\
\big[ \TL^A,\T^{\alpha}{}_{aa'} \big]&= \phantom{-} \tfrac{1}{2}(\lambda^A)^b{}_a \;\T^{\alpha}{}_{ba'},\\
\big[ \TR^A,\T^{\alpha}{}_{aa'} \big]&=  \phantom{-}\tfrac{1}{2}(\lambda^A)^{b'}{}_{a'} \;\T^{\alpha}{}_{ab'},\\
\big[ \TC^A,\TB_{\alpha}{}^{aa'} \big]&= \phantom{-}\tfrac{1}{2}(\lambda^A)^\beta{}_\alpha \;\TB_{\beta}{}^{aa'},\\
\big[ \TL^A,\TB_{\alpha}{}^{aa'} \big]&=-\tfrac{1}{2}(\lambda^A)^a{}_b \;\TB_{\alpha}{}^{ba'},\\
\big[ \TR^A,\TB_{\alpha}{}^{aa'} \big]&=-\tfrac{1}{2}(\lambda^A)^{a'}{}_{b'} \;\TB_{\alpha}{}^{ab'},
\end{align}
\begin{align}
\big[ \T^{\alpha}{}_{aa'},\T^{\beta}{}_{bb'} \big]&=-\varepsilon^{\alpha\beta\gamma}\; \varepsilon_{abc}\; \varepsilon_{a'b'c'} \;\TB_{\gamma}{}^{cc'},\\
\big[ \TB_{\alpha}{}^{aa'},\TB_{\beta}{}^{bb'} \big]&=\phantom{-}\varepsilon_{\alpha\beta\gamma}\; \varepsilon^{abc}\; \varepsilon^{a'b'c'} \;\T^{\gamma}{}_{cc'},
\end{align}
\begin{align}
\big[ \TB_{\alpha}{}^{aa'},\T^{\beta}{}_{bb'} \big]&=(\lambda^A)^\beta{}_\alpha \;\delta^a{}_b \;\delta^{a'}{}_{b'} \;\TC^A-\delta^\beta{}_\alpha \;(\lambda^A)^{a}{}_{b} \;\delta^{a'}{}_{b'} \;\TL^A-\delta^\beta{}_\alpha \;\delta^{a}{}_{b} \; (\lambda^A)^{a'}{}_{b'} \;\TR^A. \label{eq:commutation-e6-end}
\end{align}

These relation are a modified version of those from~\cite{Kephart:1981gf}. We use a conjugate embedding for the C $\SU(3)$-factor relative to those of L and R in $G_{333}$, in accordance with  Eq.~\eqref{eq:adjoint-into-trinification}. Our preferred embedding is phenomenologically more convenient and is also the more common one adopted in the  literature~\cite{Slansky:1981yr,Bajc:2013qra,Babu:2015psa,Feger:2012bs,Feger:2019tvk}, while \cite{Kephart:1981gf} implements a more symmetric embedding with the last two terms in Eq.~\eqref{eq:adjoint-into-trinification} replaced by $(\mathbf{3},\mathbf{3},\mathbf{3})\oplus (\mathbf{\bar{3}},\mathbf{\bar{3}},\mathbf{\bar{3}})$.

Note that the only objects necessary to define $\mathrm{E}_{6}$ commutation relations are $\SU(3)$-related:
\begin{enumerate}
	\item The Gell-Mann matrices $\lambda^{A}$.
	\item The $\SU(3)$ structure constants $f^{ABC}$:
		\begin{align}
		[\tfrac{1}{2}\lambda^{A},\tfrac{1}{2}\lambda^{B}]&= i f^{ABC}\,\tfrac{1}{2}\lambda^{C}.
		\end{align}
	\item The $\SU(3)$ invariant tensors $\varepsilon_{abc}$, $\varepsilon^{abc}$, $\delta^{a}{}_{b}$ (the completely anti-symmetric Levi-Civita and Kronecker delta tensors in three dimensions).
\end{enumerate}
Schematically, the commutation relations by sector can be understood most easily from Table~\ref{table:schematic-commutation-e6}.

\begin{table}[htb]
\centering
\caption{The schematic table of the $\mathrm{E}_{6}$ commutation relations by sector. \label{table:schematic-commutation-e6}}
\begin{tabular}{r|lllll}
	$[.,.]$&$T_{C}$&$T_{L}$&$T_{R}$&$t$&$\bar{t}$\\
	\hline
	$T_C$&$T_C$&0&0&$t$&$\bar{t}$\\
	$T_L$&0&$T_L$&0&$t$&$\bar{t}$\\
	$T_R$&0&0&$T_R$&$t$&$\bar{t}$\\
	$t$&$t$&$t$&$t$&$\bar{t}$&$T_{C,L,R}$\\
	$\bar{t}$&$\bar{t}$&$\bar{t}$&$\bar{t}$&$T_{C,L,R}$&$t$\\
\end{tabular}
\end{table}

For completeness, we now provide also the action of the $\mathrm{E}_{6}$ generators on the fundamental representation $\mathbf{27}$. The representation $\mathbf{27}$ decomposes under the $G_{333}$ subgroup (in our chosen embedding) as
\begin{align}
	\mathbf{27}&= (3,3,1)\oplus (1,\bar{3},3)\oplus (\bar{3},1,\bar{3})
	\equiv L^{\alpha a}\oplus M_{a}{}^{a'}\oplus N_{a'\alpha}. \label{eq:decompose-27}
\end{align}
In this $(L^{\alpha a},M_{a}{}^{a'},N_{a'\alpha})$ notation for the $\mathbf{27}$, the generators act via 

\def\LMN{\begin{pmatrix}
        L^{\alpha a},& 
        M_{a}{}^{a'},& 
        N_{a'\alpha}\\
        \end{pmatrix}
}

\begin{align}
T_C^A \; \LMN &=
    \begin{pmatrix}
        \tfrac{1}{2}(\lambda_A)^{\alpha}{}_\beta \; L^{\beta a},&
        0,&
        -\tfrac{1}{2}(\lambda^{\ast}_A)_{\alpha}{}^{\beta} \; N_{a' \beta}\\
    \end{pmatrix}, \label{eq:E6-action27-begin}\\
T_L^{A}\; \LMN &=
    \begin{pmatrix}
        \tfrac{1}{2}(\lambda_A)^{a}{}_b \; L^{\alpha b},&
        - \tfrac{1}{2}(\lambda^{\ast}_A)_a{}^{b} \; M_b{}^{a'},&
        0\\
    \end{pmatrix},\\
T_R^{A}\; \LMN &=
    \begin{pmatrix}
        0,&
        \tfrac{1}{2}(\lambda_A)^{a'}{}_{b'}\; M_{a}{}^{b'},&
        -\tfrac{1}{2}(\lambda^{\ast}_A)_{a'}{}^{b'} N_{b' \alpha} \\
    \end{pmatrix},\\
t^{\alpha}{}_{aa'}\;\LMN&=
    \begin{pmatrix}
        \varepsilon^{\alpha\beta\gamma} \;\delta^{b}{}_a \; N_{a'\gamma},&
        -\varepsilon_{abc} \;\delta^{b'}{}_{a'} \; L^{\alpha c},&
        -\varepsilon_{a'b'c'} \;\delta^\alpha{}_\beta \; M_a{}^{c'}\\
    \end{pmatrix},\\
\bar{t}_{\alpha}{}^{aa'} \; \LMN&=
    \begin{pmatrix}
        \varepsilon^{abc} \;\delta^\beta{}_\alpha \; M_{c}{}^{a'},&
        \varepsilon^{a'b'c'} \;\delta^{a}{}_{b} \; N_{c' \alpha},&
        -\varepsilon_{\alpha\beta\gamma} \;\delta^{a'}{}_{b'} \; L^{\gamma a}\\
    \end{pmatrix}. \label{eq:E6-action27-end}
\end{align}
The generator action has again been taken from~\cite{Kephart:1981gf} and modified in accordance with our preferred $G_{333}$ embedding, cf.~discussion below Eq.~\eqref{eq:commutation-e6-end}. Note that Eqs.~\eqref{eq:E6-action27-begin}--\eqref{eq:E6-action27-end} allow for an explicit construction of $\mathrm{E}_{6}$ generator matrices in the fundamental representation $\mathbf{27}$. These indeed satisfy the commutation relations of Eqs.~\eqref{eq:commutation-e6-begin}--\eqref{eq:commutation-e6-end}. The explicit form of the generators provides the computational machinery behind the explicit results in this paper.

Finally, the group $\mathrm{E}_{6}$ has its own primitive invariant tensors: $\delta^{i}{}_{j}$, $d_{ijk}$ and $d^{ijk}$, where $i,j,k$ are fundamental $\mathrm{E}_{6}$ indices and run from $1$ to $27$. Their components can be extracted from the relation
\begin{align}
\tfrac{1}{3!}\,d_{ijk}\;\mathbf{27}^{i}\,\mathbf{27}^{j}\,\mathbf{27}^{k} &= -\det\mathbf{L}+\det\mathbf{M}-\det\mathbf{N}-\mathrm{Tr}(\mathbf{LMN}),
\end{align}
where the $\mathbf{27}^{i}$ was written in terms of matrices $(\mathbf{L},\mathbf{M},\mathbf{N})$ as in Eq.~\eqref{eq:decompose-27}. The tensors $d_{ijk}$ and $d^{ijk}$ have the same numerical components, are completely symmetric, and in a suitable basis all their entries are either $\pm 1$ or $0$, see~\cite{Bajc:2013qra}. In that basis they have the normalization
\begin{align}
	d^{ikl} d_{jkl}&= 10\,\delta^{i}{}_{j}.
\end{align}

\subsection{The algebra of $\mathrm{F}_{4}$ and its embedding into $\mathrm{E}_{6}$ \label{app:f4}}

The group $\mathrm{F}_{4}$ has $52$ generators. Making use of their decomposition into irreducible representations of the $\SU(3)_C\times\SU(3)_{LR}$ maximal subgroup, see Table~\ref{table:exceptional-groups}, the $\mathrm{F}_{4}$ generators can be labeled as follows:
\begin{align}
	(\Phi_{C})^{A},\quad
	(\Phi_{LR})^{A},\quad
	\phi^{\alpha}{}_{ab},\quad
	\bar{\phi}_{\alpha}{}^{ab},
\end{align}
where symmetry is imposed in $(a,b)$ (so that it forms a $6=(3\times 3)_{s}$ of $\SU(3)_{LR}$). As usual, $A$ is the adjoint index of $\SU(3)$, and $\{\alpha,a\}$ are fundamental for $\SU(3)_{C,LR}$. The subscripts $C$ and $LR$ in the maximal subgroup $\SU(3)^{2}$ already anticipate a particular embedding into $\mathrm{E}_{6}$. We abbreviate the labels of $\mathrm{F}_{4}$ generators by sector as $\{\Phi_{C},\Phi_{LR},\phi,\bar{\phi}\}$.

One possible embedding is to define the $\mathrm{F}_{4}$ generators in terms of $\mathrm{E}_{6}$ generators as follows:
\begin{align}
\Phi_{C}^{A}&:=T_C^{A}, \label{eq:definition-gen-f4-begin}\\
\Phi_{LR}^{A}&:= T_L^{A}+T_R^{A},\\
\phi^{\alpha}{}_{ab}&:= t^{\alpha}{}_{ab}+t^{\alpha}{}_{ba},\\
\bar{\phi}_{\alpha}{}^{ab}&:= \bar{t}_{\alpha}{}^{ab}+\bar{t}_{\alpha}{}^{ba}. \label{eq:definition-gen-f4-end}
\end{align}
Notice that the set of $\mathrm{F}_{4}$ generators is obtained from $\mathrm{E}_{6}$ generators by symmetrizing with respect to LR parity from Appendix~\ref{app:discrete-symmetries}. This is a particularly simple embedding, in the context of which the $\mathfrak{f}_{4}$ Lie algebra can be understood as the maximal subalgebra  of $\mathfrak{e}_{6}$ that consists of LR-symmetric elements.

Using the commutation relations of $\mathrm{E}_{6}$ from Eqs.~\eqref{eq:commutation-e6-begin}--\eqref{eq:commutation-e6-end}, we get the following consistent set of $\mathrm{F}_{4}$ commutation relations:
\begin{align}
\big[ \Phi_{C}^A,\Phi_{LR}^B \big]&=0, 
\end{align}
\begin{align}
\big[ \Phi_{C}^A,\Phi_{C}^B \big]&=i f^{ABC} \;\Phi_{C}^C,\\
\big[ \Phi_{LR}^A,\Phi_{LR}^B \big]&=i f^{ABC} \;\Phi_{LR}^C,
\end{align}
\begin{align}
\big[ \Phi_{C}^A,\phi^{\alpha}{}_{ab} \big]&= -\tfrac{1}{2}(\lambda^A)^\alpha{}_\beta \;\phi^{\beta}{}_{ab},\\
\big[ \Phi_{LR}^A,\phi^{\alpha}{}_{ab} \big]&= \phantom{-} \tfrac{1}{2}\left((\lambda^A)^c{}_a \;\phi^{\alpha}{}_{cb}
+ (\lambda^A)^c{}_b \;\phi^{\alpha}{}_{ac}\right),\\
\big[ \Phi_{C}^A,\bar{\phi}_{\alpha}{}^{ab} \big]&= \phantom{-}\tfrac{1}{2}(\lambda^A)^\beta{}_\alpha \;\bar{\phi}_{\beta}{}^{ab},\\
\big[ \Phi_{LR}^A,\bar{\phi}_{\alpha}{}^{ab} \big]&=-\tfrac{1}{2}\left((\lambda^A)^a{}_c \;\bar{\phi}_{\alpha}{}^{cb}+(\lambda^A)^b{}_c \;\bar{\phi}_{\alpha}{}^{ac}\right),
\end{align}
\begin{align}
\big[ \phi^{\alpha}{}_{ab},\phi^{\beta}{}_{cd} \big]&=-\varepsilon^{\alpha\beta\gamma}\;
(\varepsilon_{ace}\,\varepsilon_{bdf}+\varepsilon_{bce}\,\varepsilon_{adf}) \;\bar{\phi}_{\gamma}{}^{ef},\\
\big[ \bar{\phi}_{\alpha}{}^{ab},\bar{\phi}_{\beta}{}^{cd} \big]&=+\varepsilon_{\alpha\beta\gamma}\; 
(\varepsilon^{ace}\,\varepsilon^{bdf}+\varepsilon^{bce}\,\varepsilon^{adf}) \;\phi^{\gamma}{}_{ef},
\end{align}
\begin{align}
\big[ \bar{\phi}_{\alpha}{}^{ab},\phi^{\beta}{}_{cd} \big]&=2\,(\lambda^A)^\beta{}_\alpha \;(\delta^a{}_c \;\delta^{b}{}_{d}+\delta^{a}{}_{d}\; \delta^{b}{}_{c}) \;\Phi_{C}^A
- \nonumber\\
&\quad- \delta^{\beta}{}_{\alpha}\;\left(
(\lambda^{A})^{a}{}_{c}\,\delta^{b}{}_{d}+(\lambda^{A})^{a}{}_{d}\,\delta^{b}{}_{c}
+(\lambda^{A})^{b}{}_{d}\,\delta^{a}{}_{c}+(\lambda^{A})^{b}{}_{c}\,\delta^{a}{}_{d}\right)\; \Phi_{LR}^{A}.
\end{align}
The commutation relations are consistent with the generator $\phi^\alpha{}_{ab}$ being symmetric in the indices $a$ and $b$, as they should be.

As in $\mathrm{E}_{6}$, the commutation relations of $\mathrm{F}_{4}$ were possible to write down by using only $\SU(3)$-related objects: the Gell-Mann matrices $\lambda^{A}$, the Levi-Civita tensors $\varepsilon_{abc}$ and the $\SU(3)$ structure constants $f^{ABC}$. This is the advantage offered by considering $\mathrm{F}_{4}$ in the language of its $\SU(3)^{2}$ maximal subalgebra.

Schematically, the $\mathrm{F}_{4}$ commutation relations by sector are displayed in Table~\ref{table:schematic-commutation-f4}. The generators from different sectors do not necessarily have a consistent normalization.

\begin{table}[htb]
\centering
\caption{The schematic table of the $\mathrm{F}_{4}$ commutation relations by sector.  \label{table:schematic-commutation-f4}}
\begin{tabular}{r|llll}
	$[.,.]$&$\Phi_{C}$&$\Phi_{LR}$&$\phi$&$\bar{\phi}$\\
	\hline
	$\Phi_C$&$\Phi_C$&0&$\phi$&$\bar{\phi}$\\
	$\Phi_{LR}$&0&$\Phi_{LR}$&$\phi$&$\bar{\phi}$\\
	$\phi$&$\phi$&$\phi$&$\bar{\phi}$&$\Phi_{C,LR}$\\
	$\bar{\phi}$&$\bar{\phi}$&$\bar{\phi}$&$\Phi_{C,LR}$&$\phi$\\
\end{tabular}
\end{table}

\subsection{The algebra of $\SU(3)\times\mathrm{G}_{2}$ and its embedding into $\mathrm{E}_{6}$ \label{app:su3g2}}

We label the $8$ generators of $\SU(3)$ by
\begin{align}
	T^{A}
\end{align}
and the $14$ generators of $G_{2}$ by 
\begin{align}
	G^{A},\quad g^{a}, \quad \bar{g}_{a}.
\end{align}
We used the label $A$ for the adjoint index of $\SU(3)$, and the (upper) $a$ as the fundamental index of $\SU(3)$. Furthermore, we made use of the maximal $\SU(3)$ subgroup of $\mathrm{G}_{2}$ to decompose the generators of $\mathrm{G}_{2}$, see Table~\ref{table:exceptional-groups}. In the abbreviated notation by sector, the generators of $\SU(3)\times\mathrm{G}_{2}$ are written simply as $\{T,G,g,\bar{g}\}$.

One possible embedding of $\SU(3)\times\mathrm{G}_{2}$ into $\mathrm{E}_{6}$ is to define its generators as follows: 
\begin{align}
	T^{12+}&=T_{L}^{45+} + t^{2}{}_{23}, &
	T^{45+}&= t^{2}{}_{12} - t^{2}{}_{21}, &
	T^{67+}&= T_{R}^{45+} + t^{2}{}_{32}, \label{eq:definition-gen-su3g2-begin} \\
	T^{12-}&=T_{L}^{45-} + \bar{t}_{2}{}^{23}, &
	T^{45-}&= \bar{t}_{2}{}^{12} - \bar{t}_{2}{}^{21}, &
	T^{67-}&= T_{R}^{45-} + \bar{t}_{2}{}^{32},
\end{align}
\begin{align}
	T^{3}&= \langle \tfrac{1}{2}, -\tfrac{1}{2\sqrt{3}},
		0,\tfrac{2}{\sqrt{3}},
		0,-\tfrac{1}{\sqrt{3}}		
		\rangle, \\[3pt]
	T^{8}&= \langle 
		\tfrac{\sqrt{3}}{2},-\tfrac{1}{2},0,0,0,1
	\rangle,
\end{align}
\begin{align}
	G^{12+}&= t^{1}{}_{22}, &
	G^{45+}&= t^{3}{}_{22}, &
	G^{67+}&= T_C^{45+}, \\
	G^{12-}&= \bar{t}_{1}{}^{22}, &
	G^{45-}&= \bar{t}_{3}{}^{22}, &
	G^{67-}&= T_C^{45-},
\end{align}
\begin{align}
	G^{3}&= \langle 
		-\tfrac{1}{2}, -\tfrac{1}{2\sqrt{3}}, -\tfrac{1}{2},
		\tfrac{1}{2\sqrt{3}}, -\tfrac{1}{2}, \tfrac{1}{2\sqrt{3}}	
	\rangle \\[3pt]
	G^{8}&= \langle 
		\tfrac{1}{2\sqrt{3}}, \tfrac{5}{6}, -\tfrac{1}{2\sqrt{3}}, 
		\tfrac{1}{6}, -\tfrac{1}{2\sqrt{3}}, \tfrac{1}{6}
	\rangle,
\end{align}
\begin{align}
	g^{1}&= t_L^{12+} + t_{R}^{12+} + t^{2}{}_{33}, &
	g^{2}&= t^{1}{}_{12} + t^{1}{}_{21} + \bar{t}_{3}{}^{11}, &
	g^{3}&= t^{3}{}_{12} + t^{3}{}_{21} - \bar{t}_{1}{}^{11}, \\[3pt]
	\bar{g}_{1}&= t_L^{12-} + t_{R}^{12-} + \bar{t}_{2}{}^{33}, &
	\bar{g}_{2}&= \bar{t}_{1}{}^{12} + \bar{t}_{1}{}^{21} + t^{3}{}_{11}, &
	\bar{g}_{3}&= \bar{t}_{3}{}^{12} + \bar{t}_{3}{}^{21} - t^{1}{}_{11}. \label{eq:definition-gen-su3g2-end}
\end{align}

Above, we made use of the following ``vector notation'' for the diagonal generators:
\begin{align}
\langle \alpha_{1},\alpha_{2},\alpha_{3},\alpha_{4},\alpha_{5},\alpha_{6} \rangle&:=
\alpha_{1}\;T_{C}^{3} +
\alpha_{2}\;T_{C}^{8} +
\alpha_{3}\;T_{L}^{3} +
\alpha_{4}\;T_{L}^{8} +
\alpha_{5}\;T_{R}^{3} +
\alpha_{6}\;T_{R}^{8}.
\end{align}
The diagonal generators $\{ T^{3},T^{8},G^{3},G^{8}\}$ form a set of orthogonal states in the sense of the Killing form $K(\mathbf{X},\mathbf{Y})=\mathrm{Tr}(\mathrm{ad}_\mathbf{X}\mathrm{ad}_\mathbf{Y})$, although they are not normalized in the same way. 

Another observation is that the generators of the factor $\mathrm{G}_{2}$ are symmetric under LR parity, and hence the group $\mathrm{G}_{2}$ defined here is a subgroup of the group $\mathrm{F}_{4}$, whose embedding is defined in Appendix~\ref{app:f4}. The generators $T$ of the $\SU(3)$ factor, however, are not LR-parity symmetric, ensuring that $\SU(3)\times\mathrm{G}_{2}$ is not a subgroup of $\mathrm{F}_{4}$ and can thus be a maximal subgroup of $\mathrm{E}_{6}$.

Inserting the commutation relations of $\mathrm{E}_{6}$ from Eqs.~\eqref{eq:commutation-e6-begin}--\eqref{eq:commutation-e6-end}, it is possible to work out the following consistent set of $\SU(3)\times\mathrm{G}_{2}$ commutation relations:

\begin{align}
	[T^{A},G^{A}]&=[T^{A},g^{a}]=[T^{A},\bar{g}_{a}]=0,
\end{align}
\begin{align}
	[T^{A},T^{B}]&=i f^{ABC}\;T^C,
\end{align}
\begin{align}
	[G^{A},G^{B}]&=i f^{ABC}\;G^C,
\end{align}
\begin{align}
	[G^{A},g^{a}]&= -\tfrac{1}{2}\,(\lambda^{A})^{a}{}_{b}\; g^{b},\\
	[G^{A},\bar{g}_{a}]&= +\tfrac{1}{2}\,(\lambda^{A})^{b}{}_{a}\; \bar{g}_{b}
\end{align}
\begin{align}
	[g^{a},g^{b}]&= -2\, \varepsilon^{abc} \; \bar{g}_{c},\\
	[\bar{g}_{a},\bar{g}_b]&= +2\, \varepsilon_{abc} \; g^{c},
\end{align}
\begin{align}
	[\bar{g}_{a},g^{b}]&= 3\, (\lambda^{A})^{b}{}_{a}\, G^{A}.
\end{align}
Note that the above commutation relations use $T^{A}$ and $G^{A}$ in the Hermitian basis, while  the generator definitions of Eq.~\eqref{eq:definition-gen-su3g2-begin}--\eqref{eq:definition-gen-su3g2-end} were given in a complex basis of raising and lowering operators. The conversion between them can be done via
the relations
\begin{align}
	T^{AB\pm}&=T^{A}\pm i T^{B},&
	G^{AB\pm}&=G^{A}\pm i G^{B}. 
\end{align} 
Analogous to $\mathrm{E}_{6}$ and $\mathrm{F}_{4}$, the language of the maximal subgroup $\SU(3)$ in $\mathrm{G}_{2}$ again allowed us to write down the $\mathrm{G}_{2}$ commutation relations only with $\SU(3)$-related objects. In fact, the use of $\SU(3)$-covariant indices structurally already restricts the form of the commutation relations up to overall numeric coefficients. The commutation relations also have a straightforward interpretation in this $\SU(3)$ language, for example the relation for $[G^{A},g^{a}]$ simply states that $g^{a}$ transform as a triplet under the $\SU(3)$ subgroup defined by $G^{A}$.

For better intuition, we again write the commutation relations of $\mathrm{G}_{2}$ by sector in Table~\ref{table:schematic-commutation-g2}. The generators from different sectors do not necessarily have a consistent normalization.

\begin{table}[htb]
\centering
\caption{The schematic table of the $\mathrm{G}_{2}$ commutation relations by sector. \label{table:schematic-commutation-g2}}
\begin{tabular}{r|lll}
	$[.,.]$&$G$&$g$&$\bar{g}$\\
	\hline
	$G$&$G$&$g$&$\bar{g}$\\[3pt]
	$g$&$g$&$\bar{g}$&$G$\\[3pt]
	$\bar{g}$&$\bar{g}$&$G$&$g$\\
\end{tabular}
\end{table}

\section{Discrete symmetries in $\mathrm{E}_6$ \label{app:discrete-symmetries}}

One could consider reshuffling the labels for color (C), left (L) and right (R) of the $\SU(3)$ factors in trinification $G_{333}$. Such transformations form a permutation group of $3$ objects, which we denoted by $D_{3}$ in Section~\ref{sec:maximal-subgroups}. The group $D_{3}$ is generated by 3 parities, which exchange pairs of $\SU(3)$ factors: $\mathbb{Z}_{2}^{LR}$, $\mathbb{Z}_{2}^{CL}$ and $\mathbb{Z}_{2}^{CR}$, also dubbed LR, CL and CR parity. 

The permutation group $D_{3}$ is a discrete subgroup of $\mathrm{E}_{6}$. This can be seen by explicitly constructing the parities; it turns out we can define them by
\begin{align}
    P_{LR}&:=e^{i\pi A_{LR}}\,e^{i\pi B_{LR}},&
    P_{CL}&:=e^{i\pi A_{CL}}\,e^{i\pi B_{CL}},&
    P_{CR}&:=e^{i\pi A_{CR}}\,e^{i\pi B_{CR}}, \label{eq:parities-definition-begin}
\end{align}
where the $A$- and $B$-matrices are all $\mathfrak{e}_{6}$ algebra elements: 
\begin{align}
    A_{LR}&:=\tfrac{1}{2i}(t^{1}{}_{22}-\bar{t}_{1}{}^{22})+\tfrac{1}{2i}(t^{1}{}_{11}-\bar{t}_{1}{}^{11}),&
    B_{LR}&:=\tfrac{1}{2i}(t^{1}{}_{33}-\bar{t}_{1}{}^{33})-t^{7}_{C}, \label{eq:parities-definition-LR}\\
    A_{CL}&:=\tfrac{1}{2i}(t^{2}{}_{21}-\bar{t}_{2}{}^{21})+\tfrac{1}{2i}(t^{1}{}_{11}-\bar{t}_{1}{}^{11}),&
    B_{CL}&:=\tfrac{1}{2i}(t^{3}{}_{31}-\bar{t}_{3}{}^{31})-t^{7}_{R},\\
    A_{CR}&:=\tfrac{1}{2i}(t^{2}{}_{12}-\bar{t}_{2}{}^{12})+\tfrac{1}{2i}(t^{1}{}_{11}-\bar{t}_{1}{}^{11}),&
    B_{CR}&:=\tfrac{1}{2i}(t^{3}{}_{13}-\bar{t}_{3}{}^{13})-t^{7}_{L}. \label{eq:parities-definition-end}
\end{align}

The actions of the parities on the $\mathbf{27}$ can then be conveniently specified in the language of the $(\mathbf{L},\mathbf{M},\mathbf{N})$ matrices from Eq.~\eqref{eq:decompose-27}. Using the parity definitions of Eqs.~\eqref{eq:parities-definition-begin}--\eqref{eq:parities-definition-end} and the action of $\mathrm{E}_{6}$ generators on the $\mathbf{27}$ from Eqs.~\eqref{eq:E6-action27-begin}--\eqref{eq:E6-action27-end}, we obtain
\begin{align}
	P_{LR}:&\quad (\mathbf{L},\mathbf{M},\mathbf{N})
		\mapsto 
		(\mathbf{N}^\TRANSPOSE,\mathbf{M}^\TRANSPOSE,\mathbf{L}^\TRANSPOSE),
		\label{eq:PLR-onfundamental}\\
	P_{CL}:&\quad (\mathbf{L},\mathbf{M},\mathbf{N})
		\mapsto 
		(\mathbf{L}^\TRANSPOSE,-\mathbf{N}^\TRANSPOSE,-\mathbf{M}^\TRANSPOSE),
		\label{eq:PCL-onfundamental}\\
	P_{CR}:&\quad (\mathbf{L},\mathbf{M},\mathbf{N})
		\mapsto 
		(-\mathbf{M}^\TRANSPOSE,-\mathbf{L}^\TRANSPOSE,\mathbf{N}^\TRANSPOSE).
		\label{eq:PCR-onfundamental}
\end{align}
The parities thus act as specific permutations of the states in the representation $\mathbf{27}$ (up to minus signs). The presence of minuses is due to our use of the phenomenologically more convenient embedding of trinification, see discussion below Eq.~\eqref{eq:commutation-e6-end}.

Armed with the prescription of Eqs.~\eqref{eq:PLR-onfundamental}--\eqref{eq:PCR-onfundamental}, we can easily confirm that the defined transformations are parities in the mathematical sense:
\begin{align}
    P_{LR}^2&=P_{CL}^2=P_{CR}^2=\mathbb{1}.
\end{align}
Furthermore, we define the composite transformation $P_{CLR}\equiv P_{LR}P_{CR}$; the group elements of $D_{3}$ can now be written as
\begin{align}
    D_{3}&=\{ \mathbb{1}, P_{CLR}, P_{CLR}^{2}, P_{LR}, P_{CL}, P_{RC} \}.
\end{align}
It can be checked explicitly that the set closes under multiplication, and the group structure is indeed that of a permutation group $S_{3}$ of $3$ elements. The element 
$P_{CLR}$ generates a $\mathbb{Z}_{3}$ cyclic subgroup, and can be understood intuitively to cyclically permute $C\mapsto L\mapsto R \mapsto C$. In the context of 
Section~\ref{sec:vacuum-solutions} and Eq.~\eqref{eq:R-rotation}, we can identify $P_{CLR}$ with clockwise rotations of $3$-fold symmetry: $P_{CLR}=R(4\pi/3)$ and $P_{CLR}^{2}=R(2\pi/3)$.

The action of the parities can be extended to any representation of $\mathrm{E}_{6}$ by tensoriality. We now consider how the parities act on the adjoint $\mathbf{78}$, which in tensor form has one upper and one lower index, cf.~Table~\ref{table:E6-irreps}. The action of a parity $P$ on a generator $\mathbf{T}$ (written as a matrix) is then given explicitly by 
\begin{align}
	\hat{P}(\mathbf{T})&= \mathbf{P\,T\,P}^\TRANSPOSE, \label{eq:parity-on-adjoint}
\end{align}
where the matrix $\mathbf{P}$ is constructed via Eqs.~\eqref{eq:PLR-onfundamental}--\eqref{eq:PCR-onfundamental}. This gives the explicit results of Table~\ref{table:parity-on-adjoint}. The expressions confirm the intuition behind what these parities do, i.e.~they exchange $\SU(3)$ factors (up to taking the generator in the conjugate representation). 

\begin{table}[htb]
    \centering
    \caption{The action of the $3$ parities $\mathbb{Z}_{2}^{LR}$, $\mathbb{Z}_{2}^{CL}$ on the basis elements of the adjoint representation $\mathbf{78}$ of $\mathrm{E}_{6}$ (when the adjoint representation is considered as the span of generators in the fundamental representation). \label{table:parity-on-adjoint}}
    \vspace{-0.2cm}
    \begin{tabular}{lccccc}
        \toprule
        map $P$&
        $\hat{P}(t^{A}_{C})$&
        $\hat{P}(t^{A}_{L})$&
        $\hat{P}(t^{A}_{R})$&
        $\hat{P}(t^\alpha{}_{aa'})$&
        $\hat{P}(\bar{t}_\alpha{}^{aa'})$\\
        \midrule
        $P_{LR}$&
            $-(t^{A}_{C})^\TRANSPOSE$&
            $-(t^{A}_{R})^\TRANSPOSE$&
            $-(t^{A}_{L})^\TRANSPOSE$&
            $-(t^{\alpha}{}_{a'a})^\TRANSPOSE$&
            $-(\bar{t}_{\alpha}{}^{a'a})^\TRANSPOSE$\\
        $P_{CL}$&
            $t^{A}_{L}$&
            $t^{A}_{C}$&
            $-(t^{A}_{R})^\TRANSPOSE$&
            $-(t^{a}{}_{\alpha a'})^\TRANSPOSE$&
            $-(\bar{t}_{a}{}^{\alpha a'})^\TRANSPOSE$\\
        $P_{CR}$&
            $t^{A}_{R}$&
            $-(t^{A}_{L})^\TRANSPOSE$&
            $t^{A}_{C}$&
            $-(t^{a'}{}_{a\alpha})^\TRANSPOSE$&
            $-(\bar{t}_{a'}{}^{a\alpha})^\TRANSPOSE$\\
        \bottomrule
    \end{tabular}
\end{table}

To summarize, we saw explicitly that Eqs.~\eqref{eq:parities-definition-begin}--\eqref{eq:parities-definition-end} indeed define parities forming the discrete $D_{3}$ subgroup of $\mathrm{E}_{6}$ with all intuitive properties of exchanging the $\SU(3)$ factor labels of the $G_{333}$ subgroup. 

The importance of the $D_{3}$ group is obvious from the vacuum analysis when breaking along the $G_{333}$ direction. In particular, it appears as a symmetry of the restricted potential of $G_{333}$-singlets in Eq.~\eqref{eq:singlet-potential-1} and is visually manifest in Figure~\ref{fig:restricted-potential}. Furthermore, the $G_{333}$ vacua preserve 
a $\mathbb{Z}_{2}$ remnant (one of the parities), cf.~Section~\ref{sec:vacuum-solutions}. Placing ourselves in e.g.~the LR symmetric vacuum,  the full preserved symmetry is in fact $G_{333}\rtimes \mathbb{Z}_{2}^{LR}$, as confirmed also by the LR symmetric spectrum in Table~\ref{tab:masses-vacuum-su3su3su3}. We emphasize that the product with the discrete group is semi-direct, since elements of $D_{3}$ do not commute with those of $G_{333}$, i.e.~they act on $G_{333}$ elements non-trivially in accordance with Table~\ref{table:parity-on-adjoint}.

Finally, we observe that given the definition of $P_{LR}$ in Eqs.~\eqref{eq:parities-definition-begin} and \eqref{eq:parities-definition-LR}, LR parity is also part of the $\SO(10)$ subgroup of $\mathrm{E}_{6}$ that is specified (together with an Abelian factor) in Eq.~\eqref{eq:define-rsubgroups-so10}. LR parity can thus be identified (up to phase conventions) as $D$-parity~\cite{Kibble:1982dd,Chang:1983fu,Chang:1984uy,Chang:1984qr} in the $\SO(10)$ context, and has also been referred to as such in the context of trinification originating from $\mathrm{E}_{6}$~\cite{Chakrabortty:2019fov,Dash:2020jlc}. Following these references, $D$-parity is defined as
\begin{align}
    P_{D}&=e^{i\pi T_{67}}e^{i\pi T_{23}}=-\Sigma_{67}\Sigma_{23},\label{eq:definition-D-parity}
\end{align}
where $T_{ij}\equiv \Sigma_{ij}/2=[\Gamma_{i},\Gamma_{j}]/(4i)$ are the $\SO(10)$ generators in the spinorial representation\footnote{Throughout the paper, we refer to groups with orthogonal algebras simply as $\SO(n)$, but strictly speaking we in fact always have the simply-connected variants $\mathrm{Spin}(n)$ in mind. The reader should take this into account when topological considerations or representation theory are discussed.} (the full $\mathbf{16}\oplus\mathbf{\overline{16}}$), with $\Gamma_{i}$ the gamma matrices of the $10$-dimensional Clifford algebra $\{\Gamma_{i},\Gamma_{j}\}=2\delta_{ij}$, where indices $i,j$ run from $1$ to $10$. Note that the definition assumes the embedding of the Pati-Salam group $G_{422}\equiv \SU(4)_C\times\SU(2)_L\times\SU(2)_{R}$ into $\SO(10)$ in a convention, where the indices $1\leq i,j\leq 6$ refer to the $\SO(6)\cong \SU(4)_C$ factor. 
The definition of $P_{D}$ can then be extended to all other $\SO(10)$ representations by tensoriality. Observe that the last equality in Eq.~\eqref{eq:definition-D-parity}, i.e.~the evaluation of the exponentials, is a simplification possible in $\SO(10)$ due to the Clifford algebra relation, while no such simplification is possible in the case of $\mathrm{E}_{6}$ parities of Eq.~\eqref{eq:parities-definition-begin}.

Also, as a last clarification, we do not aspire to consider any of the $\mathrm{E}_{6}$ parities as an extension of the space-time parity as done in the left-right model~\cite{Mohapatra:1974gc,Mohapatra:1974hk,Senjanovic:1975rk}.

\section{Location of singlets in the $\mathbf{650}$ \label{appendix:states-in-650}}

We first consider the fundamental representation $\mathbf{27}$, whose decomposition into trinification representations is given in Eq.~\eqref{eq:decompose-27}. In the spirit of Appendix~A in~\cite{Bajc:2013qra}, the field content of this representation is even better elucidated in terms of its SM irreducible components and the use of SM-fermion language:
\begin{align}
	\mathbf{27}&= (Q\oplus L \oplus d^{c} \oplus u^{c} \oplus e^{c}) \oplus (d'\oplus d'^{c}\oplus L'\oplus L'^{c})\oplus (\nu^{c}\oplus n), \label{eq:27-to-SM}
\end{align}
where these fields transform under the SM group as
\begin{align}
	Q&\sim (\mathbf{3},\mathbf{2},+\tfrac{1}{6}), &
	d^{c},d'^{c} &\sim (\mathbf{\bar{3}},\mathbf{1},+\tfrac{1}{3}),&
	L,L'&\sim (\mathbf{1},\mathbf{2},-\tfrac{1}{2}),&
	e^{c} &\sim (\mathbf{1},\mathbf{1},+1),\nonumber\\
	u^{c} &\sim (\mathbf{\bar{3}},\mathbf{1},-\tfrac{2}{3}),  &
	d'&\sim (\mathbf{3},\mathbf{1},-\tfrac{1}{3}),&
	L'^{c} &\sim (\mathbf{1},\mathbf{2},+\tfrac{1}{2}),&
	\nu^{c},n &\sim (\mathbf{1},\mathbf{1},0).	
\end{align}
The first grouping of representations (in terms of parentheses) in Eq.~\eqref{eq:27-to-SM} consists of all SM fermions in one generation. The second grouping 
consists of vector-like down-type quarks $d'\oplus d'^{c}$ and vector-like lepton doublets $L'\oplus L'^{c}$. The third grouping contains SM singlets: $\nu^{c}$ is a right-handed neutrino in the subpart $\mathbf{16}$ of the $\mathrm{SO}(10)$ subgroup, and $n$ is a $\SO(10)$ singlet (denoted by $s$ in~\cite{Bajc:2013qra}). 
Writing in terms of $\SU(3)_C\times\UU_{EM}$ components of the EW broken phase, we use the notation
\begin{align}
	Q&=(u,d), & 
	L&=(\nu,e), &
	L'&=(\nu',e') &
	L'^{c} &= (\nu'^c,e'^c).
\end{align}
We suppressed all color indices in the notation. Collecting all the states together, the $\mathbf{27}$ is explicitly written in the following basis:
\begin{align}
\mathbf{e}_{i}=\{ u,\; d,\; d',\; u^{c},\; d^{c},\; d'^{c},\; e,\; e',\; e^{c},\; e'^{c},\; \nu,\; \nu',\; \nu'^{c},\; \nu^{c},\; n
\}. \label{eq:basis-27}
\end{align}
For the sake of being fully explicit, we insert this basis into the matrices $(L^{\alpha a},M_{a}{}^{a'},N_{a'\alpha})$ of Eq.~\eqref{eq:decompose-27} in the following way:
\begin{align}
	\mathbf{L}&=
		\begin{pmatrix} 
			u_{1} & d_{1} & d'_{1}\\ 
			u_{2} & d_{2} & d'_{2}\\ 
			u_{3} & d_{3} & d'_{3}\\ 
		\end{pmatrix}, &
	\mathbf{M}&=
		\begin{pmatrix}
			\nu'^{c} & e' & e \\
			e'^{c} & -\nu' & -\nu\\
			e^{c} & \nu^{c} & n\\
		\end{pmatrix}, &
	\mathbf{N}&= 
		\begin{pmatrix}
			u^{c}_{1} & u^{c}_{2} & u^{c}_{3} \\
			-d^{c}_{1} & -d^{c}_{2} & -d^{c}_{3} \\
			d'^{c}_{1} & d'^{c}_{2} & d'^{c}_{3} \\
		\end{pmatrix}.
\end{align} 
The quarks were written with explicit color indices in their subscripts, and the
presence of minuses is part of our conventions for greater convenience. The two minuses in $\mathbf{M}$ are present due to this representation transforming as a $\mathbf{\bar{3}}$ under $\SU(3)_L$, and hence the lepton doublets $L$ and $L'$ are embedded as a $\mathbf{\bar{2}}$ of $\SU(2)_L$; the minuses ensure the definitions align with those in the SM. Similarly, the minuses in $\mathbf{N}$ indicate the $\SU(2)_R$ quark doublet $(u^{c},d^{c})$ is embedded as a $\mathbf{\bar{2}}$.

We are now ready to consider the representation $\mathbf{650}$. Since its components $X^{i}{}_{j}$ are written with one upper and one lower fundamental index, cf.~Table~\ref{table:E6-irreps}, its basis is a $650$-dimensional subspace spanned by $\mathbf{e}_{i}\otimes \mathbf{e}^{\ast j}\equiv \mathbf{e}_{i}\mathbf{e}^{\ast j}$. It is in this ``$\mathbf{e}\mathbf{e}^{\ast}$'' basis that we specify some the states of interest in the $\mathbf{650}\equiv\mathbf{X}$.

\def\PaU{uu^{\ast}}
\def\PaD{dd^{\ast}}
\def\PaDP{d'd'^{\ast}}
\def\PaNPC{\nu'^{c}\nu'^{c\ast}}
\def\PaEP{e'e'^{\ast}}
\def\PaE{ee^{\ast}}
\def\PaEPC{e'^c e'^{c\ast}}
\def\PaNP{\nu'\nu'^{\ast}}
\def\PaN{\nu\nu^{\ast}}
\def\PaEC{e^{c}e^{c\ast}}
\def\PaNC{\nu^{c}\nu^{c\ast}}
\def\PaS{nn^{\ast}}
\def\PaUC{u^{c}u^{c\ast}}
\def\PaDC{d^{c}d^{c\ast}}
\def\PaDPC{d'^{c}d'^{c\ast}}

The SM singlets listed in Eq.~\eqref{eq:650-SM-singlets} and Table~\ref{table:650-singlet-content} have the following specific form in $\mathbf{X}$:
\begingroup
\allowdisplaybreaks
\begin{align}
s&\simeq \tfrac{1}{6\sqrt{3}}\big(
 (\PaU +\PaD +\PaDP) \nonumber\\ 
 &\qquad -2(\PaNPC +\PaEP +\PaE
  +\PaEPC +\PaNP +\PaN
  +\PaEC  +\PaNC +\PaS) \nonumber\\
 &\qquad +(\PaUC +\PaDC +\PaDPC) \big) \label{eq:650-SM-singlets-begin}\\
a&\simeq \tfrac{1}{6} \big(
(\PaU+\PaD+\PaDP)-(\PaUC+\PaDC+\PaDPC) \big)
\\
x_1&\simeq \tfrac{1}{6\sqrt{2}} \big(
2(\PaUC+\PaNPC+\PaEPC +\PaEC) \nonumber\\
 &\qquad -(\PaDC +\PaDPC +\PaEP +\PaE +\PaNP +\PaN +\PaNC +\PaS)
\big)\\
x_2&\simeq \tfrac{1}{2\sqrt{6}} \big(
(\PaDC + \PaEP + \PaNP + \PaNC)
- (\PaDPC + \PaE + \PaN + \PaS)
\big)\\
X_3&\simeq \tfrac{1}{2\sqrt{3}}(
e' e^{\ast} + \nu' \nu^{\ast} + \nu^{c} n^{\ast} + d'^{c} d^{c\ast}
)\\
y_1&\simeq \tfrac{1}{6\sqrt{2}} \big(
(2\PaNPC -\PaEP -\PaE)
  +(2 \PaEPC -\PaNP -\PaN)
  -2(2\PaEC  - \PaNC -\PaS)
\big)\\
y_2&\simeq \tfrac{1}{2\sqrt{6}}\big(
 -(\PaEP -\PaE)
  -(\PaNP -\PaN)
  +2(\PaNC -\PaS)
\big)\\
Y_3&\simeq \tfrac{1}{2\sqrt{3}} ( e' e^{\ast} + \nu' \nu^{\ast}-2 \nu^{c} n^{\ast})\\
z&\simeq \tfrac{1}{6\sqrt{2}} \big(
(\PaU +\PaD + \PaNPC +\PaEP +\PaE
  +\PaEPC +\PaNP +\PaN) \nonumber\\
 &\qquad -2 (\PaDP + \PaEC  +\PaNC +\PaS)
\big). \label{eq:650-SM-singlets-end}
\end{align}
\endgroup
The conjugates $X_3^{\ast}$ and $Y_{3}^{\ast}$ are obtained by exchanging the order in each term and complex conjugating each fermion label (equivalent to Hermitian conjugation in $\mathbf{X}$). Also, the states $X_{3}$ and $Y_{3}$ are the only SM-singlet states with off-diagonal entries in $\mathbf{X}$. These states are normalized to $1/2$ for real fields and $1$ for complex fields:
\begin{align}
X^{i}{}_{j}\,X^\ast{}_{i}{}^{j}&= \tfrac{1}{2}(s^2+a^2+x_{1}^{2}+x_{2}^{2}+y_{1}^{2}+y_{2}^{2}+z^{2})+ |X_{3}|^{2}+|Y_{3}|^{2}.
\end{align}
As always, the overall sign for each real field and phase for a complex field is conventional.

We emphasize again that color indices were suppressed in our notation; the quark terms should thus be taken as a sum over all colors, e.g.~$uu^{\ast}\equiv u^{a}u^{\ast}{}_{a}$, where $a$ is a color index going from $1$ to $3$. 

Since these states in Eqs.~\eqref{eq:650-SM-singlets-begin}--\eqref{eq:650-SM-singlets-end} are SM singlets, they could alternatively be written in the language of the EW unbroken phase with weak indices suppressed, e.g.
\begin{align}
z&\simeq \tfrac{1}{6\sqrt{2}} \big( (QQ^\ast + LL^\ast+L'L'^\ast+L'^{c}L'^{c\ast}) 
 	-2 (\PaDP + \PaEC  +\PaNC +\PaS)
\big).
\end{align}

Finally, we specify the singlets of the exceptional cases. For the case of the $\mathrm{F}_{4}$, with its embedding specified in Eqs.~\eqref{eq:definition-gen-f4-begin}--\eqref{eq:definition-gen-f4-end}, the singlet state $\tilde{s}$ has the direction
\begin{align}
	\tilde{s}_{\mathrm{F}_{4}} &
	\simeq \frac{1}{6\,\sqrt{39}}\left(
		\sum_{i=1}^{27} \mathbf{e}_{i}\,\mathbf{e}^{\ast i}
		- 9 \left( ss^\ast+\nu'\nu'^\ast+\nu'^{c}\nu'^{c\ast} \right)
		- 9 \left( \nu'^{c}s^\ast -\nu' s^\ast - \nu'\nu'^{c\ast} + h.c.
		\right)
	\right).
\end{align}
For the other exceptional case $\SU(3)\times\mathrm{G}_{2}$ with the embedding given in Eqs.~\eqref{eq:definition-gen-su3g2-begin}--\eqref{eq:definition-gen-su3g2-end}, the singlet has the direction 
\begin{align}
	\tilde{s}_{\SU(3)\times\mathrm{G}_{2}} &
	\simeq \frac{1}{12\,\sqrt{21}} \;\bigg(
		4\,\sum_{i=1}^{27} \mathbf{e}_{i}\,\mathbf{e}^{\ast i}
	- 9 \left(
		u_{2}u_{2}^\ast + u^{c}_{2}u^{c\ast}_{2}
		+ ee^\ast + e^{c}e^{c\ast}
		+ \nu'\nu'^\ast + \nu'^{c} \nu'^{c\ast}
	\right) -  \nonumber\\[4pt]
	&\quad 
	- 18 \left(ss^\ast+d'_{2}d'^{\ast}_{2}+ d'^{c}_{2}d'^{c\ast}_{2} \right)
	- 9 \left(u_{2}e^\ast - \nu'\nu'^{c\ast} + u_{2}^{c}e^{c\ast} + h.c. \right)
	\bigg).
\end{align}

Notice that the singlets in the exceptional cases cannot be written in terms of 
SM representations, i.e.~$\tilde{s}_{\mathrm{F}_{4}}$ explicitly breaks $\SU(2)_L$ while $\tilde{s}_{\SU(3)\times\mathrm{G}_{2}}$ breaks both $\SU(2)_L$ and $\SU(3)_C$. This is not surprising, since the exceptional cases do not have a SM group as a subgroup, and thus their singlets are not SM singlets.



\input{bib1.bbl}

\end{document}

%% file: bib1.bbl
\providecommand{\href}[2]{#2}\begingroup\raggedright\endgroup